\documentclass[british,nofootinbib, 11pt,a4paper]{revtex4-2}
\usepackage[T1]{fontenc}
\usepackage[latin9]{inputenc}
\usepackage{mathrsfs}
\usepackage{mathtools}
\usepackage{amsmath}
\usepackage{amsthm}
\usepackage{amssymb}
\theoremstyle{remark}
\newtheorem*{rem*}{\protect\remarkname}
\theoremstyle{plain}
\newtheorem{thm}{\protect\theoremname}
\theoremstyle{definition}
\newtheorem{example}[thm]{\protect\examplename}
\usepackage{babel}
\providecommand{\examplename}{Example}
\providecommand{\remarkname}{Remark}
\providecommand{\theoremname}{Theorem}

\begin{document}
\title{Partial Differential Equations for MHV Celestial Amplitudes in Liouville
Theory}
\author{Igor Mol}
\affiliation{State University of Campinas (Unicamp)}
\email{igormol@ime.unicamp.br}

\begin{abstract}
In this note, we continue our study of Liouville theory and celestial
amplitudes by deriving a set of partial differential equations governing
the $n$-point MHV celestial amplitudes for gluons and gravitons,
parametrised by the Liouville coupling constant $b$. These equations
provide a systematic framework for computing the perturbative expansion
in $b$ of the celestial amplitudes, which are known to reproduce
the tree-level MHV $n$-point functions for pure Yang-Mills and Einstein
gravity in the semiclassical $b\rightarrow0$ limit. We demonstrate
that the $\mathcal{O}(b^{2})$ corrections are logarithmic for both
gluons and gravitons. Additionally, we compute the correction to the
celestial operator product expansion (OPE) parametrised by $b^{2}$
and show that it is isomorphic to the one-loop correction to the celestial
OPEs of gluons and gravitons. We propose that celestial Liouville
theory, extended beyond its semiclassical limit, encodes the one-loop
corrections to Yang-Mills theory and perturbative gravity in flat
spacetime.
\end{abstract}
\maketitle
\tableofcontents{}

\section{Introduction}

An important challenge in the bottom-up approach to celestial holography
is the construction of simplified models for celestial conformal field
theory (CFT) that capture key features of realistic theories in asymptotically
flat spacetime backgrounds. Recent studies by \citet{stieberger2023celestial,stieberger2023yang,giribet2024remarks,melton2024celestial,mol2024holographic}
have shown that the tree-level MHV celestial amplitudes for both gluons
and gravitons can be expressed as the semiclassical limit, $b\rightarrow0$,
of correlation functions of dressed Liouville vertex operators on
the celestial sphere. A natural question arises: Can one extend this
correspondence beyond the semiclassical regime of Liouville theory
to describe other sectors of Yang-Mills and Einstein theories? In
this note, we continue the investigation of the connection between
Liouville theory and celestial amplitudes, and suggest that the answer
is affirmative. Specifically, we present evidence that Liouville theory
beyond the semiclassical limit encodes the one-loop corrections to
Yang-Mills theory and perturbative gravity in flat space.

The perspective adopted in these notes is as follows. We begin by
reviewing how the scattering amplitudes for tree-level MHV gravitons
and gluons can be expressed as the semiclassical limit of correlation
functions in Liouville theory. In particular, we analyse in detail
how the BGK formula, proposed by \citet{berends1988relations}, for
tree-level MHV $n$-graviton amplitudes admits an operator factorisation
especially amenable to Mellin transformation, making it particularly
interesting in the context of celestial amplitudes. Building upon
this observation, in Sections \ref{sec:Gravitons} and \ref{sec:Differential-Equations-for-Gluons},
we propose an ansatz for what we term the \emph{celestial Liouville
amplitudes}, which extends the original celestial amplitudes beyond
the semiclassical limit $b\rightarrow0$ of the Liouville sector.
Subsequently, using the Ward identities for Liouville theory, we derive
a system of partial differential equations that characterises the
celestial Liouville amplitudes for both gluons and gravitons, parametrised
by the Liouville coupling constant, $b$. These equations are particularly
well-suited for perturbative analysis, thus providing a systematic
framework to compute corrections to celestial amplitudes.

The analysis of correlation functions in field theories is greatly
facilitated by the study of differential and functional equations,
as demonstrated in the seminal works of \citet{knizhnik1984current,belavin1984infinite,dolan2001conformal,zinn2005renormalization}.
In the framework of celestial holography, a significant system of
partial differential equations for MHV graviton amplitudes was derived
by \citet{banerjee2021mhv,banerjee2021subsubleading}. This line of
inquiry was expanded upon by \citet{pasterski2021celestial,hu2023celestial,hu2021celestial}.
Additionally, in the context of the Carrollian approach, we mention
the recent contribution by \citet{ruzziconi2024differential}.

In Sections \ref{sec:Celestial-OPE-Gravitons} and \ref{sec:Celestial-OPE-Gluons},
we shall study the leading-order perturbative corrections to the celestial
Liouville amplitudes for gluons and gravitons, derived from our system
of partial differential equations. As we will demonstrate, these corrections
are logarithmic for both pure Yang-Mills theory and Einstein gravity.
Subsequently, we employ the collinear limit approach within the splitting
function formalism, developed by \citet{bern1999multi,white2011factorization,akhoury2011collinear},
and first applied to celestial holography by \citet{pate2021celestial},
to compute the leading-order corrections to the celestial OPEs for
the Liouville amplitudes. These corrections will be identified with
the one-loop corrections to Yang-Mills theory and Einstein gravity,
as studied by \citet{bhardwaj2022loop,bhardwaj2024celestial,bittleston2023associativity,krishna2024celestial,adamo2023all,banerjee2023infinite,banerjee2023all,banerjee2024all}.
Finally, in Section \ref{sec:Discussion}, we summarise our results
and outline potential directions for future research.

\section{Differential Equations for MHV Gravitons\label{sec:Gravitons}}

In this section, we derive a system of partial differential equations
characterising the celestial amplitudes for the scattering of MHV
gravitons, parametrised by the Liouville coupling constant $b$. Our
approach is based on the observation, following the works of \citet{stieberger2023celestial,stieberger2023yang,melton2024celestial,giribet2024remarks,mol2024holographic},
that the celestial amplitude for \emph{tree-level }MHV graviton scattering
can be expressed as the semiclassical limit $b\rightarrow0$ of Liouville
correlation functions. Using this result as an ansatz for the celestial
Liouville amplitude, we employ the Ward identities for Liouville correlation
functions to derive a system of coupled partial differential equations
for these amplitudes. This system can be solved perturbatively in
$b^{2}$, providing a framework for systematically incorporating corrections
from the Liouville sector into the celestial CFT. In Section \ref{sec:Celestial-OPE-Gravitons},
we compute the leading-order correction in $b^{2}$, which is then
used to determine the deformation of the celestial OPE for gravitons
using the collinear limit approach with the splitting function formalism,
as discussed in the context of celestial holography by \citet{pate2021celestial}.
We show that the resulting OPE algebra is isomorphic to the celestial
OPE for gravitons incorporating the one-loop correction to Einstein
gravity as studied by \citet{bhardwaj2022loop,bhardwaj2024celestial,bittleston2023associativity,krishna2024celestial,adamo2023all}.

\subsection{Differential Operators for Graviton Amplitudes\label{subsec:Differential-Operators-for}}

In this subsection, our objective is to find a factorisation of $M_{n}$
using differential operators, allowing us to express it in a form
that includes a term which, after applying the Mellin transform, can
be identified with the semiclassical limit of Liouville theory correlators. 

We begin by recalling the formula derived by \citet{berends1988relations}
for the tree-level scattering of MHV $n$-gravitons with\footnote{A similar analysis applies straightforwardly for $n=3$ and $n=4$,
so we may proceed without loss of generality by considering the case
$n\geq5$.} $n\geq5$:
\begin{align}
M_{n} & =\delta_{P}\frac{\left\langle 12\right\rangle ^{8}}{\left\langle 12\right\rangle \left\langle 23\right\rangle ...\left\langle n1\right\rangle }\frac{1}{\left\langle 1,n\right\rangle \left\langle 1,n-1\right\rangle \left\langle n-1,n\right\rangle }\prod^{n-2}_{k=2}\frac{[k\big|p_{k+1}+...+p_{n-1}\big|n\rangle}{\left\langle kn\right\rangle }\label{eq:Berends}\\
 & +\mathcal{P}\left(2,...,n-2\right)
\end{align}
where $\delta_{P}\coloneqq\delta^{\left(4\right)}\left(P\right)$
is the energy-momentum conservation delta function, with $P=p_{1}+...+p_{n}$.

Next, we introduce a doublet of chiral primary free fermions $\chi\left(z\right)$
and $\chi^{\dagger}\left(z\right)$ on $\mathbf{CP}^{1}$, defined
by the mode expansions:
\begin{equation}
\chi\left(z\right)\coloneqq\sum^{\infty}_{n=0}b_{n}z^{n},\,\,\,\chi^{\dagger}\left(z\right)\coloneqq\sum^{\infty}_{n=0}b^{\dagger}_{n}z^{-n-1},
\end{equation}
where $b_{n}$ and $b^{\dagger}_{n}$ are fermionic annihilation and
creation operators satisfying the anti-commutation relations $\{b_{m},b^{\dagger}_{n}\}=\delta_{mn}$
for all $m,n\geq0$, and act on the vacuum $\big|0\rangle$ as $b_{n}\big|0\rangle=0$
for each non-negative integer $n$. The two-point function for this
doublet is given by:
\begin{equation}
\langle\chi\left(z_{1}\right)\chi^{\dagger}\left(z_{2}\right)\rangle\coloneqq\langle0\big|\chi\left(z_{1}\right)\chi^{\dagger}\left(z_{2}\right)\big|0\rangle=\frac{1}{z_{12}}.
\end{equation}

To facilitate our analysis, we define another pair of fermions on
the space of two-component spinors $\mathbf{C}^{2}$. Let $\mu^{A}\coloneqq\left(\xi,\zeta\right)\in\mathbf{C}^{2}$
be a two-component spinor with $\zeta\neq0$, and let $z=\xi/\zeta\in\mathbf{CP}^{1}$
be its local coordinate on an open neighbourhood $\mathcal{U}\subset\mathbf{CP}^{1}$
on the complex projective line. We define the fermions $\hat{\chi}\left(\mu\right)$
and $\hat{\chi}^{\dagger}\left(\mu\right)$ on $\mathbf{C}^{2}$ as:

\begin{equation}
\hat{\chi}\left(\mu\right)\coloneqq\frac{1}{\zeta}\chi\left(z\right),\,\,\,\hat{\chi}^{\dagger}\left(\mu\right)\coloneqq\frac{1}{\zeta}\chi^{\dagger}\left(z\right),
\end{equation}
on the neighbourhood $\mathcal{U}$. If $\zeta=0$, one can choose
another coordinate neighbourhood $\left(z',\mathcal{U}'\right)$ in
the atlas of $\mathbf{CP}^{1}$ and use $z'=\zeta/\xi$. The two-point
function for these fermions is then:
\begin{equation}
\langle\hat{\chi}\left(\nu_{j}\right)\hat{\chi}^{\dagger}\left(\nu_{i}\right)\rangle=\frac{1}{\left\langle ij\right\rangle }.
\end{equation}

For each pair of two-component spinors $\nu^{A}_{i}=\left(z_{i},1\right)^{T}$
and $\bar{\nu}^{\dot{A}}_{i}=\left(\bar{z}_{i},1\right)^{T}$, parametrising
the null four-momentum $p^{\mu}_{i}\left(z_{i},\bar{z}_{i}\right)=\omega_{i}(\sigma^{\mu})_{A\dot{A}}\nu^{A}_{i}\bar{\nu}^{\dot{A}}_{i}$,
define the operators:
\begin{equation}
\mathcal{Q}_{i}\coloneqq e^{ip_{i}\cdot x}\hat{\chi}^{\dagger}\left(\nu_{i}\right)\hat{\chi}\left(\nu_{i}\right)=\frac{e^{i\omega_{i}q\left(z_{i},\bar{z}_{i}\right)\cdot x}}{\omega_{i}}\chi^{\dagger}\left(z_{i}\right)\chi\left(z_{i}\right),\label{eq:Q-Operator}
\end{equation}
\begin{equation}
\mathcal{P}_{i}\coloneqq-i\frac{\bar{\nu}^{\dot{A}}_{i}\lambda^{A}}{\left\langle \nu_{i},\lambda\right\rangle }\frac{\partial}{\partial x^{A\dot{A}}}e^{i\omega_{i}q\left(z_{i},\bar{z}_{i}\right)\cdot x},\label{eq:P-Operator}
\end{equation}
where $\lambda^{A}\in\mathbf{C}^{2}$ is an auxiliary two-component
spinor with local coordinate $\lambda\in\mathbf{C}$, over which we
will integrate in our final expression. Consequently, letting $\big|\lambda\rangle\coloneqq\chi^{\dagger}\left(\lambda\right)\big|0\rangle$,
it can be shown (cf. \citet{mol2024holographic}) that Eq. (\ref{eq:Berends})
may be rewritten as:
\begin{align}
M_{n} & =\left(\prod^{n}_{k=1}\omega^{\alpha_{k}}_{k}\right)\frac{z^{8}_{12}}{z_{12}z_{23}...z_{n1}}\int\frac{d^{4}x}{\left(2\pi\right)^{4}}\int\frac{d^{2}\lambda}{\pi}\langle\lambda\big|\mathcal{Q}_{1}\left(\prod^{n-2}_{r=2}\mathcal{P}_{r}\right)\mathcal{Q}_{n-1}\mathcal{Q}_{n}\bar{\partial}\big|\lambda\rangle\label{eq:Berends-1}\\
 & +\mathcal{P}\left(2,...,n-2\right),
\end{align}
where $\alpha_{1}=\alpha_{2}=3$ and $\alpha_{i}=-1$ for all $3\leq i\leq n$.

\subsection{Celestial Amplitude for Gravitons\label{subsec:Celestial-Amplitude-for}}

In this subsection, we will provide a detailed computation of the
Mellin transform of Eq. (\ref{eq:Berends-1}) to derive the celestial
amplitude $\widehat{M}_{n}$ for the tree-level scattering amplitude
of MHV gravitons. This will be done using the celestial holography
dictionary (see \citet{pasterski2021lectures,raclariu2021lectures}),
according to which:
\begin{equation}
\widehat{M}_{n}\left(\Delta_{1},z_{1},\bar{z}_{1};...;\Delta_{n},z_{n},\bar{z}_{n}\right)=\prod^{n}_{k=1}\int^{\infty}_{0}d\omega_{k}\omega^{\Delta_{k}-1}_{k}e^{-\epsilon\omega_{k}}M_{n}\left(\omega_{1},z_{1},\bar{z}_{1};...;\omega_{n},z_{n},\bar{z}_{n}\right).\label{eq:Celestial-Amplitude}
\end{equation}

We define the $\epsilon$-regulated Mellin-transformed operators corresponding
to $\mathcal{Q}$ and $\mathcal{P}$, with conformal weight $\Delta$
and frequency power $\alpha\in\mathbb{Z}$, as follows:
\begin{equation}
\widehat{\mathcal{Q}}_{i}\coloneqq\int^{\infty}_{0}d\omega\omega^{\left(\Delta_{i}+\alpha_{i}\right)-1}e^{-\epsilon\omega}\mathcal{Q}_{i},\label{eq:Mellin-Transformed-Operator-Q}
\end{equation}
\begin{equation}
\widehat{\mathcal{P}}_{i}\coloneqq\int^{\infty}_{0}d\omega\omega^{\left(\Delta_{i}+\alpha_{i}\right)-1}e^{-\epsilon\omega}\mathcal{P}_{i}.\label{eq:Mellin-Transformed-Operator-P}
\end{equation}
Using Eqs. (\ref{eq:Q-Operator}, \ref{eq:P-Operator}), and defining
the parameters $\sigma_{i}$ by:
\begin{equation}
\sigma_{i}\coloneqq\frac{1}{2}\left(\Delta_{i}+\alpha_{i}-1\right)\,\,\,\text{for \ensuremath{i\in\left\{ 1,n-1,n\right\} },}\label{eq:Sigma}
\end{equation}
\begin{equation}
\sigma_{i}\coloneqq\frac{1}{2}\left(\Delta_{i}+\alpha_{i}\right)\,\,\,\text{for \ensuremath{2\leq i\leq n-2}},\label{eq:Sigma-1}
\end{equation}
the Mellin-transformed operators $\widehat{\text{\ensuremath{\mathcal{Q}}}}$
and $\widehat{\mathcal{P}}$ becomes:
\begin{equation}
\widehat{\mathcal{Q}}_{i}=\phi_{2\sigma_{i}}\left(z_{i},\bar{z}_{i},x\right)\chi^{\dagger}\left(z\right)\chi\left(z\right),\label{eq:Mellin-Transformed-Q-Operator}
\end{equation}
\begin{equation}
\widehat{\mathcal{P}}_{i}=-i\frac{\bar{\nu}^{\dot{A}}\lambda^{A}}{\left\langle \nu,\lambda\right\rangle }\frac{\partial}{\partial x^{A\dot{A}}}\phi_{2\sigma_{i}}\left(z_{i},\bar{z}_{i},x\right),\label{eq:Mellin-Transformed-P-Operator}
\end{equation}
where $\phi_{h}\left(z,\bar{z},x\right)$ is the scalar conformal
primary wavefunction with scaling dimension $h$ (cf. \citet{pasterski2017conformal,pasterski2017flat}),
given by:
\begin{equation}
\phi_{h}\left(z,\bar{z},x\right)=\frac{\Gamma\left(h\right)}{\left(\varepsilon-iq\left(z,\bar{z}\right)\cdot x\right)^{h}}.
\end{equation}
It is useful to introduce the operator $\mathsf{P}_{i}$, acting on
the scalar conformal primaries $\phi_{2\sigma_{j}}\left(z_{j},\bar{z}_{j},x\right)$,
by shifting the conformal weights by a half unit, defined as:
\begin{equation}
\mathsf{P}_{i}\phi_{2\sigma_{j}}\left(z_{j},\bar{z}_{j}\right)\coloneqq\delta_{ij}\phi_{2\left(\sigma_{j}+1/2\right)}\left(z_{j},\bar{z}_{j},x\right).\label{eq:Shifting-Operator}
\end{equation}

Hence, using Eqs. (\ref{eq:Berends-1}, \ref{eq:Celestial-Amplitude})
and Eqs. (\ref{eq:Mellin-Transformed-Q-Operator}, \ref{eq:Mellin-Transformed-P-Operator}),
the celestial amplitude can be expressed as:
\begin{equation}
\widehat{M}_{n}=\frac{z^{8}_{12}}{z_{12}z_{23}...z_{n1}}\int\frac{d^{4}x}{\left(2\pi\right)^{4}}\int\frac{d^{2}\lambda}{\pi}\langle\lambda\big|\widehat{\mathcal{Q}}_{1}\left(\prod^{n-2}_{r=2}\widehat{\mathcal{P}}_{r}\right)\mathcal{\widehat{\mathcal{Q}}}_{n-1}\mathcal{\widehat{\mathcal{Q}}}_{n}\bar{\partial}\big|\lambda\rangle+\mathcal{P}\left(2,...,n-2\right).\label{eq:Celestial-Amplitude-1}
\end{equation}

The aforementioned expression for $\widehat{M}_{n}$ is somewhat inadequate
for our objectives. Specifically, we aim to express the celestial
amplitude as correlators of Liouville vertex operators, which necessitates
factoring the integration over the measure $d^{4}x$ outside the inner
product $\langle\lambda\big|...\bar{\partial}\big|\lambda\rangle$.
To achieve this, we adopt the following strategy.

Let us define the new operators:
\begin{equation}
\mathsf{A}_{i}\coloneqq\exp\left(q\left(z,\bar{z}\right)\cdot y\,\mathsf{P}_{i}\right)\chi^{\dagger}\left(z\right)\chi\left(z\right),\label{eq:A-Operator}
\end{equation}
\begin{equation}
\mathsf{B}_{i}\coloneqq\frac{\bar{\nu}^{\dot{A}}_{i}\lambda^{A}}{\left\langle \nu_{i},\lambda\right\rangle }\frac{\partial}{\partial y^{A\dot{A}}}\exp\left(q\left(z,\bar{z}\right)\cdot y\,\mathsf{P}_{i}\right).\label{eq:B-Operator}
\end{equation}
Here, the four-vector $y^{\mu}$ should be interpreted as a continuous
label for these operators rather than spacetime coordinates. Thus,
from the relation:
\begin{equation}
-i\frac{\partial}{\partial x^{A\dot{A}}}\phi_{2\sigma_{j}}\left(z_{j},\bar{z}_{j},x\right)=\int dy\delta\left(y\right)\frac{\partial}{\partial y^{A\dot{A}}}\exp\left(q\left(z,\bar{z}\right)\cdot y\,\mathsf{P}_{j}\right)\phi_{2\sigma_{j}}\left(z_{j},\bar{z}_{j},x\right),
\end{equation}
it can be shown, using induction (cf. \citet{mol2024holographic}),
that:
\begin{equation}
\langle\lambda\big|\widehat{\mathcal{Q}}_{1}\left(\prod^{n-2}_{i=2}\widehat{\mathcal{P}}_{i}\right)\mathcal{\widehat{\mathcal{Q}}}_{n-1}\mathcal{\widehat{\mathcal{Q}}}_{n}\bar{\partial}\big|\lambda\rangle=\int d\mu\left(y\right)\langle\lambda\big|\mathsf{A}_{1}\left(\prod^{n-2}_{i=2}\mathsf{B}_{i}\right)\mathsf{A}_{n-1}\mathsf{A}_{n}\bar{\partial}\big|\lambda\rangle\prod^{n}_{j=1}\phi_{2\sigma_{j}}\left(z_{j},\bar{z}_{j},x\right),
\end{equation}
where, for brevity, we have introduced the symbol $\int dy\delta\left(y\right)\coloneqq\int d^{4}y\delta^{\left(4\right)}\left(y\right)$.

Consequently, the celestial amplitude for the scattering of $n$-gravitons
can be expressed as:
\begin{align}
\widehat{M}_{n} & =\frac{z^{8}_{12}}{z_{12}z_{23}...z_{n1}}\int\frac{d^{2}\lambda}{\pi}\int dy\delta\left(y\right)\langle\lambda\big|\mathsf{A}_{1}\left(\prod^{n-2}_{i=2}\mathsf{B}_{i}\right)\mathsf{A}_{n-1}\mathsf{A}_{n}\bar{\partial}\big|\lambda\rangle\\
 & \int\frac{d^{4}x}{\left(2\pi\right)^{4}}\prod^{n}_{j=1}\phi_{2\sigma_{j}}\left(z_{j},\bar{z}_{j},x\right)+\mathcal{P}\left(2,...,n-2\right).\label{eq:Celestial-1}
\end{align}

\begin{rem*}
The spacetime integral has been successfully factored out from the
inner product $\langle\lambda\big|...\bar{\partial}\big|\lambda\rangle$,
as desired.
\end{rem*}
As demonstrated in the Appendix \ref{sec:Operator-Factorisation},
the action of the operators $\mathsf{B}_{2}$, ..., $\mathsf{B}_{n-2}$
on $\mathsf{A}_{n-1}\mathsf{A}_{n}$ can be expanded as follows:
\begin{equation}
\left(\prod^{n-2}_{i=2}\mathsf{B}_{i}\right)\mathsf{A}_{n-1}\mathsf{A}_{n}=\mathcal{X}\prod^{n}_{r=2}e^{q\left(z_{r},\bar{z}_{r}\right)\cdot y\,\mathsf{P}_{r}}\left(\prod^{n-2}_{i=2}\sum^{n}_{j_{i}=i+1}\right)\prod^{n-2}_{k=2}\frac{\left(\bar{z}_{k}-\bar{z}_{j_{k}}\right)\left(z_{j_{k}}-\lambda\right)}{z_{k}-\lambda}\mathsf{P}_{j_{k}},\label{eq:Operator-Expansion}
\end{equation}
where $\mathcal{X}\coloneqq\prod^{n}_{i=n-1}\chi^{\dagger}\left(z_{i}\right)\chi\left(z_{i}\right)$.
Hence, defining the operator:
\begin{equation}
\mathscr{L}^{\left(n\right)}_{\sigma_{1},...,\sigma_{n}}\left(z_{1},\bar{z}_{1};...,z_{n},\bar{z}_{n}\right)\coloneqq\int\frac{d^{2}\lambda}{\pi}\int dy\delta\left(y\right)\langle\lambda\big|\mathsf{A}_{1}\left(\prod^{n-2}_{r=2}\mathsf{B}_{r}\right)\mathsf{A}_{n-1}\mathsf{A}_{n}\bar{\partial}\big|\lambda\rangle,
\end{equation}
it follows from Eq. (\ref{eq:Operator-Expansion}) that $\mathscr{L}^{\left(n\right)}_{\sigma_{1},...,\sigma_{n}}$
is linear operator with respect to the variables $\sigma_{1},...,\sigma_{n}$,
which can be expanded in terms of the operators $\mathsf{P}_{i}$
defined in Eq. (\ref{eq:Shifting-Operator}),
\begin{equation}
\mathscr{L}^{\left(n\right)}_{\sigma_{1},...,\sigma_{n}}=\frac{1}{z_{n\left(n-1\right)}z_{\left(n-1\right)1}z_{1n}}\left(\prod^{n-2}_{i=2}\sum^{n}_{j_{i}=i+1}\right)\prod^{n-2}_{k=2}\frac{\left(\bar{z}_{k}-\bar{z}_{j_{k}}\right)\left(z_{j_{k}}-z_{n}\right)}{z_{k}-z_{n}}\mathsf{P}_{j_{k}}\label{eq:Linear-Operator}
\end{equation}

Substituting Eq. (\ref{eq:Linear-Operator}) into the formula for
the celestial amplitude $\widehat{M}_{n}$ given by Eq. (\ref{eq:Celestial-1}),
we finally arrive at:
\begin{equation}
\widehat{M}_{n}=\frac{z^{8}_{12}}{z_{12}z_{23}...z_{n1}}\mathscr{L}^{\left(n\right)}_{\sigma_{1},...,\sigma_{n}}\left[\int\frac{d^{4}x}{\left(2\pi\right)^{4}}\prod^{n}_{k=1}\phi_{2\sigma_{k}}\left(z_{k},\bar{z}_{k},x\right)\right]+\mathcal{P}_{2,...,n-2}.\,\,\,\label{eq:Step}
\end{equation}

We now proceed to solve the spacetime integral:
\begin{equation}
\int\frac{d^{4}x}{\left(2\pi\right)^{4}}\prod^{n}_{k=1}\phi_{2\sigma_{k}}\left(z_{k},\bar{z}_{k},x\right),
\end{equation}
using the celestial leaf amplitude approach developed by \citet{melton2023celestial}.
The strategy involves analytically continuing the integral from Minkowski
spacetime $\mathbf{R}^{1}_{3}$ to the ultra-hyperbolic Klein space
$\mathbf{R}^{2}_{2}$, and subsequently expressing the latter as a
weighted integral over the leaves of the $AdS_{3}/\mathbf{Z}$ hyperbolic
foliation of $\mathbf{R}^{2}_{2}$. This procedure results in a product
involving the conformal weight delta function:
\begin{equation}
\tilde{\delta}\left(\sigma\right)\coloneqq\frac{1}{8\pi^{3}}\delta\left(4-2\sum^{n}_{i=1}\sigma_{i}\right),
\end{equation}
and a contact Feynman-Witten diagram for massless scalars propagating
on $AdS_{3}/\mathbf{Z}$, such that:
\begin{equation}
\int\frac{d^{4}x}{\left(2\pi\right)^{4}}\prod^{n}_{i=1}\phi_{2\sigma_{i}}\left(z_{i},\bar{z}_{i},x\right)=\tilde{\delta}\left(\sigma\right)\int_{AdS_{3}/\mathbf{Z}}d^{3}\hat{x}\prod^{n}_{i=1}\phi_{2\sigma_{i}}\left(z_{i},\bar{z}_{i},\hat{x}\right)+\left(\left(\bar{z}\rightarrow-\bar{z}\right)\right).\label{eq:Step-1}
\end{equation}
Here, the notation $\left(\left(\bar{z}\rightarrow-\bar{z}\right)\right)$
indicates that the first term is repeated with $\bar{z}_{i}$ replaced
by $-\bar{z}_{i}$ for all $1\leq i\leq n$.

Next, let $\phi$ be a Liouville field on $\mathbf{CP}^{1}$ subject
to the boundary conditions outlined by \citet[Ch. 2]{harlow2011analytic},
and let $\mathcal{V}_{\alpha}=:e^{2\alpha\phi}:$ be the Liouville
vertex operator with momentum $\alpha$. As demonstrated in the Appendix
of \citet{melton2024celestial}, the integral in Eq. (\ref{eq:Step-1})
can be expressed as the semiclassical limit $b\rightarrow0^{+}$ of
the correlation function of the vertex operators:
\begin{equation}
\int_{AdS_{3}/\mathbf{Z}}d^{3}\hat{x}\prod^{n}_{i=1}\phi_{2\sigma_{i}}\left(z_{i},\bar{z}_{i},\hat{x}\right)=\frac{1}{\mathcal{N}_{\mu,b}}\lim_{b\rightarrow0^{+}}\left\langle \prod^{n}_{i=1}\Gamma\left(2\sigma_{i}\right)\mathcal{V}_{b\sigma_{i}}\left(z_{i},\bar{z}_{i}\right)\right\rangle ,\label{eq:Step-2}
\end{equation}

Consequently, from Eqs. (\ref{eq:Step}, \ref{eq:Step-1}, \ref{eq:Step-2})
and the analytic continuation of the identity (cf. \citet{donnay2020asymptotic}):
\begin{equation}
\delta\left(\text{Im}\beta\right)=\frac{1}{2\pi}\int^{\infty}_{0}d\tau\tau^{-\beta-1},
\end{equation}
it follows that $\widehat{M}_{n}$ can be expressed as:
\begin{align}
\widehat{M}_{n} & =\frac{1}{\left(2\pi\right)^{4}\mathcal{N}_{\mu,b}}\frac{z^{8}_{12}}{z_{12}z_{23}...z_{n1}}\lim_{b\rightarrow0^{+}}\mathscr{L}^{\left(n\right)}_{\sigma_{1},...,\sigma_{n}}\int^{\infty}_{0}d\tau\tau^{3}\left\langle \prod^{n}_{i=1}e^{-2\sigma_{i}\log\tau}\Gamma\left(2\sigma_{i}\right)\mathcal{V}_{b\sigma_{i}}\left(z_{i},\bar{z}_{i}\right)\right\rangle \label{eq:Final}\\
 & +\mathcal{P}_{2,...,n-2}+\left(\left(\bar{z}\rightarrow-\bar{z}\right)\right).
\end{align}

\begin{rem*}
It is important to note that the Liouville vertex operators $\mathcal{V}_{b\sigma_{k}}\left(z_{k},\bar{z}_{k}\right)$
inherit, from the scalar conformal primary wavefunctions $\phi_{2\sigma_{j}}\left(z_{j},\bar{z}_{j},x\right)$,
an action from the operators $\mathsf{P}_{i}$, such that:
\begin{equation}
\mathsf{P}_{j}\mathcal{V}_{b\sigma_{k}}\left(z_{k},\bar{z}_{k}\right)=\delta_{jk}\mathcal{V}_{b\left(\sigma_{k}+1/2\right)}\left(z_{k},\bar{z}_{k}\right).
\end{equation}
This action arises from the dependence of the Liouville momentum on
$\sigma_{k}$. (If $\alpha$ is a Liouville momentum independent of
the parameters $\sigma_{k}$ for $1\leq k\leq n$, then we would have
$\mathsf{P}_{j}\mathcal{V}_{\alpha}\left(z,\bar{z}\right)=\mathcal{V}_{\alpha}\left(z,\bar{z}\right)\mathsf{P}_{j}$.)
Thus, the operator $\mathscr{L}^{\left(n\right)}_{\sigma_{1},...,\sigma_{n}}$
acts non-trivially on the Liouville correlation function: 
\[
\left\langle \prod^{n}_{k=1}\mathcal{V}_{b\sigma_{k}}\left(z_{k},\bar{z}_{k}\right)\right\rangle ,
\]
influencing the order of terms appearing in Eq. (\ref{eq:Final}).
\end{rem*}
Finally, we take the crucial step of our proposal, where we define
the $n$-point \emph{celestial Liouville amplitude} for MHV gravitons.
Let $J^{a}$ be a level-one $SO\left(2N\right)$ Wess-Zumino-Novikov-Witten
($WZNW_{1}$) current algebra on the celestial sphere. The asymptotic
limit of the Knizhnik-Zamolodchikov (KZ) equation for $SO\left(2N\right)$,
as formulated by \citet{knizhnik1984current}, determines the large-$N$
limit of the correlation functions\footnote{This result can be understood as follows. Recall that the dual Coxter
number for the Lie algebra $\mathfrak{so}_{2N}=D_{N}$ is $2N-2$,
and the KZ equation for the correlation function of the $SO\left(2N\right)$
$WZNW_{1}$ currents $J^{a}$ is given by:
\begin{equation}
\left(\frac{\partial}{\partial z_{i}}+\frac{1}{2N-1}\sum_{j\neq i}\frac{T^{a}_{i}T^{a}_{j}}{z_{i}-z_{j}}\right)\left\langle J\left(z_{1}\right)...J\left(z_{n}\right)\right\rangle =0.
\end{equation}
Next, decompose $\left\langle J\left(z_{1}\right)...J\left(z_{n}\right)\right\rangle $
into the canonical $SO\left(2N\right)$ invariants $I_{1}=\delta_{m_{1},m_{2}}\delta_{m_{3},m_{4}}$
and $I_{2}=\delta_{m_{1},m_{3}}\delta_{m_{2},m_{4}}$. Retain only
the terms $T^{a}_{i}T^{a}_{j}I_{1}$ and $T^{a}_{i}T^{a}_{j}I_{2}$
proportional to $N$, and then take the $N\rightarrow0$ limit of
the KZ equation to find the asymptotic solution given by Eq. (\ref{eq:WZNW}).}:
\begin{equation}
\left\langle \prod^{n}_{i=1}J^{a_{i}}\left(z_{i}\right)\right\rangle \Bigg|_{\mathrm{\text{single-trace}}}\sim\text{tr}\left(\prod^{n}_{i=1}T^{a_{i}}\right)\mathcal{C}_{n}(z_{1},\dots,z_{n})\,\left(N\rightarrow\infty\right),\label{eq:WZNW}
\end{equation}
where $T^{a}$ are the generators of the adjoint representation of
$SO\left(2N\right)$, normalised by the condition $\text{tr}\left(T^{a}T^{b}\right)\coloneqq2\delta^{ab}$,
and
\begin{equation}
\mathcal{C}_{n}(z_{1},\dots,z_{n})\coloneqq\frac{1}{z_{12}z_{23}...z_{n1}}\,.
\end{equation}
 The \emph{celestial Liouville amplitude} in the large-$N$ limit
is then given by:
\begin{align}
\widehat{\mathcal{M}}_{n,b} & \coloneqq\frac{z^{8}_{12}}{\left(2\pi\right)^{4}\mathcal{N}_{\mu,b}}\mathcal{C}_{n}(z_{1},\dots,z_{n})\mathscr{L}^{\left(n\right)}_{\sigma_{1},...,\sigma_{n}}\label{eq:Celestial-Liouville-Amplitude}\\
 & \int^{\infty}_{0}d\tau\tau^{3}\left\langle \prod^{n}_{i=1}e^{-2\sigma_{i}\log\tau}\Gamma\left(2\sigma_{i}\right)\mathcal{V}_{b\sigma_{i}}\left(z_{i},\bar{z}_{i}\right)\right\rangle +\mathcal{P}_{2,...,n-2}+\left(\left(\bar{z}\rightarrow-\bar{z}\right)\right).
\end{align}

Thus, the celestial Liouville amplitude $\widehat{\mathcal{M}}_{n}$
is the analytic extension of the tree-level MHV scattering amplitude
$\widehat{M}_{n}$ in Mellin space to non-zero values of the Liouville
coupling parameter $b$. Our objective now is to derive a partial
differential equation that characterises $\widehat{\mathcal{M}}_{n}$,
allowing for a systematic perturbative expansion to obtain corrections
to the original celestial amplitude, parametrised by $b$. Using this
equation, we will compute the leading $\mathcal{O}(b^{2})$ correction
to $\widehat{\mathcal{M}}_{n}$, enabling us to determine the celestial
OPE governing these amplitudes. This result will be identified with
the one-loop correction to Einstein gravity studied by \citet{adamo2023all,krishna2024celestial,bhardwaj2024celestial,bittleston2023associativity}.

It proves useful to rewrite Eq. (\ref{eq:Final}) in terms of a generating
functional. Let us define:
\begin{align}
\widehat{\mathcal{F}}_{n}\left(\sigma,z,\bar{z}\right) & \coloneqq\frac{1}{\left(2\pi\right)^{4}\mathcal{N}_{\mu,b}}\frac{z^{8}_{12}}{z_{12}z_{23}...z_{n1}}\int^{\infty}_{0}d\tau^{3}\left\langle \prod^{n}_{i=1}e^{-2\sigma_{i}\log\tau}\Gamma\left(2\sigma_{i}\right)\mathcal{V}_{b\sigma_{i}}\left(z_{i},\bar{z}_{i}\right)\right\rangle \label{eq:Generating-Function}\\
 & +\mathcal{P}_{2,...,n-2}+\left(\left(\bar{z}\rightarrow-\bar{z}\right)\right),
\end{align}
so that Eq. (\ref{eq:Celestial-Liouville-Amplitude}) can be written
compactly as:
\begin{equation}
\widehat{\mathcal{M}}_{n,b}=\mathscr{L}^{\left(n\right)}_{\sigma_{1},...,\sigma_{n}}\widehat{\mathcal{F}}_{n}\left(\sigma,z,\bar{z}\right).\label{eq:Celestial-Liouville-Amplitude-1}
\end{equation}
The inverse Mellin transform of the generating function is defined
as:
\begin{equation}
\mathcal{F}_{n}\left(\omega,z,\bar{z}\right)\coloneqq\prod^{n}_{a=1}\int^{c+i\infty}_{c-i\infty}\frac{d\Delta_{a}}{2\pi i}\omega^{-\Delta_{a}}_{a}\widehat{\mathcal{F}}_{n}\left(\sigma,z,\bar{z}\right),\label{eq:Generating-Function-1}
\end{equation}
where the relations between $\sigma_{1},...,\sigma_{n}$ and $\Delta_{1},...,\Delta_{n}$
are provided by Eqs. (\ref{eq:Sigma}, \ref{eq:Sigma-1}).

\subsection{Partial Differential Equations for MHV Gravitons\label{subsec:Partial-Differential-Equation}}

In this subsection, we will use the Ward identities for the correlation
functions of the Liouville vertex operators:
\begin{equation}
\sum^{n}_{i=1}\left(z_{i}\frac{\partial}{\partial z_{i}}+d_{i}\right)\left\langle \prod^{n}_{j=1}\mathcal{V}_{b\sigma_{j}}\left(z_{j},\bar{z}_{j}\right)\right\rangle =0,\label{eq:Ward-1}
\end{equation}
to derive a system of partial differential equations characterising
the celestial Liouville amplitudes. Here, $d_{i}$ represents the
conformal weight of the vertex operator $\mathcal{V}_{b\sigma_{i}}$
in the Liouville sector. Recalling that the conformal weight of a
vertex operator $\mathcal{V}_{\alpha}$ with Liouville momentum $\alpha$
is $\alpha\left(Q-\alpha\right)$ (cf. \citet{harlow2011analytic}),
we find:
\begin{equation}
d_{i}=\sigma_{i}+b^{2}\sigma_{i}-b^{2}\sigma^{2}_{i}.\label{eq:Conformal-Weights}
\end{equation}

Similarly, by applying Euler theorem for homogeneous functions and
recalling the expansion of $\mathscr{L}^{\left(n\right)}_{\sigma_{1},...,\sigma_{n}}$
given in Eq. (\ref{eq:Linear-Operator}), we obtain the following
relation:
\begin{equation}
\sum^{n}_{i=1}z_{i}\frac{\partial}{\partial z_{i}}\left(\frac{z^{8}_{12}}{z_{12}z_{23}...z_{n1}}\mathscr{L}^{\left(n\right)}_{\sigma_{1},...,\sigma_{n}}\right)=\left(5-n\right)\frac{z^{8}_{12}}{z_{12}z_{23}...z_{n1}}\mathscr{L}^{\left(n\right)}_{\sigma_{1},...,\sigma_{n}}.\label{eq:Identity}
\end{equation}

In our derivation, we employ the inverse Mellin transform of the operator
$\mathscr{L}^{\left(n\right)}_{\sigma_{1},...,\sigma_{n}}$. The first
step is to examine how the operator $\mathsf{P}_{j}$, defined in
Eq. (\ref{eq:Shifting-Operator}), behaves under the inverse Mellin
transform. Let $\hat{f}\left(\Delta_{a}\right)$ represent the Mellin
transform of $f\left(\omega_{a}\right)$. Then:
\begin{equation}
\int^{c+i\infty}_{c-i\infty}\frac{d\Delta_{a}}{2\pi i}\omega^{-\Delta_{a}}_{a}\mathsf{P}_{j_{b}}\hat{f}\left(\Delta_{a}\right)=\omega^{\delta_{aj_{b}}}_{a}f\left(\omega_{a}\right).
\end{equation}
This generalises to functions of several variables:
\begin{align}
 & \left(\prod^{n}_{a=1}\int^{c+i\infty}_{c-i\infty}\frac{d\Delta_{a}}{2\pi i}\omega^{-\Delta_{a}}_{a}\right)\prod^{n-2}_{b=2}\mathsf{P}_{j_{b}}\hat{f}\left(\Delta_{1},...,\Delta_{n}\right)\\
 & =\prod^{n}_{a=1}\omega^{\delta_{aj_{2}}+\delta_{aj_{3}}+...+\delta_{aj_{n-2}}}_{a}f\left(\omega_{1},...,\omega_{n}\right)\\
 & =\prod^{n}_{a=1}\prod^{n-2}_{b=2}\omega^{\delta_{aj_{b}}}_{a}f\left(\omega_{1},...,\omega_{n}\right).
\end{align}
Thus, using Eq. (\ref{eq:Linear-Operator}), we find:
\begin{align}
 & \left(\prod^{n}_{a=1}\int^{c+i\infty}_{c-i\infty}\frac{d\Delta_{a}}{2\pi i}\omega^{-\Delta_{a}}_{a}\right)\mathscr{L}^{\left(n\right)}_{\sigma_{1},...,\sigma_{n}}\hat{f}\left(\Delta_{1},...,\Delta_{n}\right)\\
 & =\frac{1}{z_{n\left(n-1\right)}z_{\left(n-1\right)1}z_{1n}}\left(\prod^{n-2}_{i=2}\sum^{n}_{j_{i}=i+1}\right)\prod^{n-2}_{k=2}\frac{\left(\bar{z}_{k}-\bar{z}_{j_{k}}\right)\left(z_{j_{k}}-z_{n}\right)}{z_{k}-z_{n}}\prod^{n}_{a=1}\omega^{\delta_{aj_{k}}}_{a}f\left(\omega_{1},...,\omega_{n}\right)\\
 & =\widetilde{\mathscr{L}}^{\left(n\right)}_{\omega_{1},...,\omega_{n}}f\left(\omega_{1},...,\omega_{n}\right),
\end{align}
where:
\begin{equation}
\widetilde{\mathscr{L}}^{\left(n\right)}_{\omega_{1},...,\omega_{n}}\left(z,\bar{z}\right)\coloneqq\frac{1}{z_{n\left(n-1\right)}z_{\left(n-1\right)1}z_{1n}}\left(\prod^{n-2}_{i=2}\sum^{n}_{j_{i}=i+1}\right)\prod^{n-2}_{k=2}\frac{\left(\bar{z}_{k}-\bar{z}_{j_{k}}\right)\left(z_{j_{k}}-z_{n}\right)}{z_{k}-z_{n}}\prod^{n}_{a=1}\omega^{\delta_{aj_{k}}}_{a},\label{eq:Inverse-Mellin-Transformed-Operator}
\end{equation}
is the inverse Mellin-transform of the operator defined in Eq. (\ref{eq:Linear-Operator}).

We now turn to the derivation of the linear differential operator
that arises from the inverse Mellin transform of the sum of conformal
weights $\sum_{i}d_{i}$ found in Eq. (\ref{eq:Conformal-Weights}).
Let $c_{n}$ and $\beta^{\left(n\right)}_{i}$ be constants such that:
\begin{equation}
\sum^{n}_{i=1}d_{i}=c_{n}+\frac{1}{2}\sum^{n}_{i=1}\beta^{\left(n\right)}_{i}\Delta_{i}-\frac{b^{2}}{4}\sum^{n}_{i=1}\Delta^{2}_{i}.\label{eq:Sigma-2}
\end{equation}
Using the identities:
\begin{equation}
\omega\frac{\partial}{\partial\omega}f\left(\omega\right)=-\int^{c+i\infty}_{c-i\infty}\frac{d\Delta}{2\pi i}\Delta\omega^{-\Delta}\hat{f}\left(\Delta\right),
\end{equation}
and:
\begin{equation}
\left(\omega\frac{\partial}{\partial\omega}\right)^{2}f\left(\omega\right)=\int^{c+i\infty}_{c-i\infty}\frac{d\Delta}{2\pi i}\Delta^{2}\omega^{-\Delta}\hat{f}\left(\Delta\right),
\end{equation}
we find that:
\begin{equation}
\left(\prod^{n}_{a=1}\int^{c+i\infty}_{c-i\infty}\frac{d\Delta_{a}}{2\pi i}\omega^{-\Delta_{a}}_{a}\right)\sum^{n}_{i=1}d_{i}\hat{f}\left(\Delta_{1},...,\Delta_{n}\right)=\mathcal{D}_{\left(n\right)}f\left(\omega_{1},...,\omega_{n}\right),
\end{equation}
where:
\begin{equation}
\mathcal{D}_{\left(n\right)}\coloneqq c_{n}-\frac{1}{2}\sum^{n}_{i=1}\beta^{\left(n\right)}_{i}\omega_{i}\frac{\partial}{\partial\omega_{i}}-\frac{b^{2}}{4}\sum^{n}_{i=1}\left(\omega_{i}\frac{\partial}{\partial\omega_{i}}\right)^{2},\label{eq:Differential-Operator}
\end{equation}
is a second-order linear differential operator with respect to the
frequencies $\omega_{1},...,\omega_{n}$.

Finally, by applying the linear differential operator $\sum_{i}z_{i}\partial/\partial z_{i}$
to the celestial Liouville amplitude $\widehat{\mathcal{M}}_{n}$,
as defined in Eqs. (\ref{eq:Celestial-Liouville-Amplitude}, \ref{eq:Generating-Function},
\ref{eq:Celestial-Liouville-Amplitude-1}), and employing the Ward
identity for the Liouville correlation functions (Eq. (\ref{eq:Ward-1})),
along with the homogeneous function theorem (Eq.(\ref{eq:Identity})),
we derive the following partial differential equation for $\widehat{\mathcal{M}}_{n}$:
\begin{equation}
\sum^{n}_{i=1}\left(z_{i}\frac{\partial}{\partial z_{i}}+n-5\right)\widehat{\mathcal{M}}_{n}+\mathscr{L}^{\left(n\right)}_{\sigma_{1},...,\sigma_{n}}\sum^{n}_{i=1}d_{i}\widehat{\mathcal{F}}_{n}=0.
\end{equation}
By taking the inverse Mellin transform and utilising Eqs. (\ref{eq:Inverse-Mellin-Transformed-Operator},
\ref{eq:Differential-Operator}), we arrive at:
\begin{equation}
\sum^{n}_{i=1}\left(z_{i}\frac{\partial}{\partial z_{i}}+n-5\right)\mathcal{M}_{n}+\widetilde{\mathscr{L}}^{\left(n\right)}_{\omega_{1},...,\omega_{n}}\mathcal{D}_{\left(n\right)}\mathcal{F}_{n}=0.\label{eq:PDE-1}
\end{equation}

Following analogous reasoning for Eq. (\ref{eq:Generating-Function}),
combined with the Ward identity in Eq. (\ref{eq:Ward-1}), and invoking
Euler theorem for homogeneous functions:
\begin{equation}
\sum^{n}_{i=1}\left(\frac{z^{8}_{12}}{z_{12}z_{23}...z_{n1}}\right)=\left(8-n\right)\frac{z^{8}_{12}}{z_{12}z_{23}...z_{n1}},
\end{equation}
we find:
\begin{equation}
\sum^{n}_{i=1}\left(z_{i}\frac{\partial}{\partial z_{i}}+n-8\right)\mathcal{F}_{n}+\mathcal{D}_{\left(n\right)}\mathcal{F}_{n}=0.\label{eq:PDE-2}
\end{equation}

Therefore, Eqs. (\ref{eq:PDE-1}, \ref{eq:PDE-2}) establish the system
of partial differential equations that govern the celestial Liouville
amplitude for MHV gravitons:
\begin{equation}
\begin{cases}
\sum^{n}_{i=1}\left(z_{i}\frac{\partial}{\partial z_{i}}+n-5\right)\mathcal{M}_{n}+\widetilde{\mathscr{L}}^{\left(n\right)}_{\omega_{1},...,\omega_{n}}\mathcal{D}_{\left(n\right)}\mathcal{F}_{n}=0,\\
\sum^{n}_{i=1}\left(z_{i}\frac{\partial}{\partial z_{i}}+n-8\right)\mathcal{F}_{n}+\mathcal{D}_{\left(n\right)}\mathcal{F}_{n}=0.
\end{cases}\label{eq:System}
\end{equation}

\begin{example}
Consider the scattering of $n=3$ MHV gravitons. From Eqs. (\ref{eq:Sigma},
\ref{eq:Sigma-1}, \ref{eq:Inverse-Mellin-Transformed-Operator}),
we find:
\begin{equation}
\sum^{n}_{i=1}d_{i}=1-2b^{2}+\frac{1}{2}\left(1-b^{2}\right)\left(\Delta_{1}+\Delta_{2}\right)+\frac{1}{2}\left(1+3b^{2}\right)\Delta_{3}-\frac{b^{2}}{4}\sum^{n}_{i=1}\Delta^{2}_{i}.
\end{equation}
The differential operator from Eq. (\ref{eq:Differential-Operator})
becomes:
\begin{equation}
\mathcal{D}_{\left(3\right)}=1-2b^{2}-\frac{1}{2}\left(1-b^{2}\right)\left(\omega_{1}\frac{\partial}{\partial\omega_{1}}+\omega_{2}\frac{\partial}{\partial\omega_{2}}\right)-\frac{1}{2}\left(1+3b^{2}\right)\omega_{3}\frac{\partial}{\partial\omega_{3}}-\frac{b^{2}}{4}\sum^{3}_{i=1}\left(\omega_{i}\frac{\partial}{\partial\omega_{i}}\right)^{2}.
\end{equation}
The inverse Mellin-transformed operator $\widetilde{\mathscr{L}}^{\left(3\right)}_{\omega_{1},\omega_{2},\omega_{3}}$
from Eq. (\ref{eq:Inverse-Mellin-Transformed-Operator}) simplifies
to:
\begin{equation}
\widetilde{\mathscr{L}}^{\left(3\right)}_{\omega_{1},\omega_{2},\omega_{3}}=\frac{1}{z_{12}z_{23}z_{31}}.
\end{equation}
Thus, the system of differential equations in Eqs. (\ref{eq:PDE-1},
\ref{eq:PDE-2}) reduces to a single partial differential equation:
\begin{align}
 & \left(\sum^{3}_{i=1}z_{i}\frac{\partial}{\partial z_{i}}-2\right)\mathcal{M}_{3}+\Bigg[1-2b^{2}-\frac{1}{2}\left(1-b^{2}\right)\left(\omega_{1}\frac{\partial}{\partial\omega_{1}}+\omega_{2}\frac{\partial}{\partial\omega_{2}}\right)\label{eq:M3}\\
 & -\frac{1}{2}\left(1+3b^{2}\right)\omega_{3}\frac{\partial}{\partial\omega_{3}}-\frac{b^{2}}{4}\sum^{3}_{i=1}\left(\omega_{i}\frac{\partial}{\partial\omega_{i}}\right)^{2}\Bigg]\mathcal{M}_{3}=0.
\end{align}
We now work perturbatively for
\begin{equation}
0<b^{2}\ll1.\label{eq:Limit}
\end{equation}
and retain terms through the first non-trivial order in $b^{2}$.
In this approximation we discard terms of order $b^{4}$ and higher,
so that Eq. (\ref{eq:M3}) simplifies to:
\begin{equation}
\left(\sum^{3}_{i=1}z_{i}\frac{\partial}{\partial z_{i}}-2\right)\mathcal{M}_{3}+\left[1-\frac{1}{2}\sum^{3}_{i=1}\omega_{i}\frac{\partial}{\partial\omega_{i}}-\frac{b^{2}}{4}\sum^{3}_{i=1}\left(\omega_{i}\frac{\partial}{\partial\omega_{i}}\right)^{2}\right]\mathcal{M}_{3}=0.
\end{equation}
\end{example}

\begin{example}
Consider the scattering of $n=4$ MHV gravitons. The sum of the conformal
weights associated with the Liouville vertex operators is given by
Eqs. (\ref{eq:Sigma}, \ref{eq:Sigma-1}, \ref{eq:Inverse-Mellin-Transformed-Operator}),
leading to:
\begin{align}
\sum^{4}_{i=1}d_{i} & =\frac{1}{2}\left(1-4b^{2}\right)+\frac{1}{2}\left(1-b^{2}\right)\Delta_{1}+\frac{1}{2}\left(1-2b^{2}\right)\Delta_{2}\\
 & +\frac{1}{2}\left(1+3b^{2}\right)\left(\Delta_{3}+\Delta_{4}\right)-\frac{b^{2}}{4}\sum^{4}_{i=1}\Delta^{2}_{i}.
\end{align}
The linear differential operator acting on the frequencies, as derived
from Eq. (\ref{eq:Differential-Operator}), takes the form:
\begin{align}
\mathcal{D}_{\left(4\right)} & =\frac{1}{2}\left(1-4b^{2}\right)-\frac{1}{2}\left(1-b^{2}\right)\omega_{1}\frac{\partial}{\partial\omega_{1}}-\frac{1}{2}\left(1-2b^{2}\right)\omega_{2}\frac{\partial}{\partial\omega_{2}}\\
 & -\frac{1}{2}\left(1+3b^{2}\right)\left(\omega_{3}\frac{\partial}{\partial\omega_{3}}+\omega_{4}\frac{\partial}{\partial\omega_{4}}\right)-\frac{b^{2}}{4}\sum^{4}_{i=1}\left(\omega_{i}\frac{\partial}{\partial\omega_{i}}\right)^{2}.
\end{align}
The operator from Eq. (\ref{eq:Inverse-Mellin-Transformed-Operator})
simplifies to:
\begin{equation}
\widetilde{\mathscr{L}}^{\left(4\right)}_{\omega_{1},...,\omega_{4}}=\frac{1}{z_{14}z_{43}z_{31}}\frac{\bar{z}_{23}z_{34}}{z_{24}}\omega_{3}.
\end{equation}
Therefore, the system of partial differential equations in Eq. (\ref{eq:System})
becomes:

\begin{equation}
\begin{aligned} & \left(\sum^{4}_{i=1}z_{i}\frac{\partial}{\partial z_{i}}-1\right)\mathcal{M}_{4}+\frac{1}{z_{14}z_{43}z_{31}}\frac{\bar{z}_{23}z_{34}}{z_{24}}\omega_{3}\Bigg[\frac{1}{2}\left(1-4b^{2}\right)-\frac{1}{2}\left(1-b^{2}\right)\omega_{1}\frac{\partial}{\partial\omega_{1}}\\
 & -\frac{1}{2}(1-2b^{2})\omega_{2}\frac{\partial}{\partial\omega_{2}}-\frac{1}{2}\left(1+3b^{2}\right)\left(\omega_{3}\frac{\partial}{\partial\omega_{3}}+\omega_{4}\frac{\partial}{\partial\omega_{4}}\right)-\frac{b^{2}}{4}\sum^{4}_{i=1}\left(\omega_{i}\frac{\partial}{\partial\omega_{i}}\right)^{2}\Bigg]\mathcal{F}_{4}=0,
\end{aligned}
\end{equation}
and:

\begin{equation}
\begin{aligned} & \left(\sum^{4}_{i=1}z_{i}\frac{\partial}{\partial z_{i}}-1\right)\mathcal{F}_{4}+\omega_{3}\Bigg[\frac{1}{2}\left(1-4b^{2}\right)-\frac{1}{2}\left(1-b^{2}\right)\omega_{1}\frac{\partial}{\partial\omega_{1}}-\frac{1}{2}(1-2b^{2})\omega_{2}\frac{\partial}{\partial\omega_{2}}\\
 & -\frac{1}{2}\left(1+3b^{2}\right)\left(\omega_{3}\frac{\partial}{\partial\omega_{3}}+\omega_{4}\frac{\partial}{\partial\omega_{4}}\right)-\frac{b^{2}}{4}\sum^{4}_{i=1}\left(\omega_{i}\frac{\partial}{\partial\omega_{i}}\right)^{2}\Bigg]\mathcal{F}_{4}=0.
\end{aligned}
\end{equation}
In the regime introduced in Eq. (\ref{eq:Limit}), these equations
simplify to:
\begin{equation}
\left(\sum^{4}_{i=1}z_{i}\frac{\partial}{\partial z_{i}}-1\right)\mathcal{M}_{4}+\frac{1}{2}\frac{1}{z_{14}z_{43}z_{31}}\frac{\bar{z}_{23}z_{34}}{z_{24}}\omega_{3}\Bigg[1-\sum^{4}_{i=1}\omega_{i}\frac{\partial}{\partial\omega_{i}}-\frac{b^{2}}{2}\sum^{4}_{i=1}\left(\omega_{i}\frac{\partial}{\partial\omega_{i}}\right)^{2}\Bigg]\mathcal{F}_{4}=0,
\end{equation}
and:
\begin{equation}
\left(\sum^{4}_{i=1}z_{i}\frac{\partial}{\partial z_{i}}-1\right)\mathcal{F}_{4}+\frac{1}{2}\Bigg[1-\sum^{4}_{i=1}\omega_{i}\frac{\partial}{\partial\omega_{i}}-\frac{b^{2}}{2}\sum^{4}_{i=1}\left(\omega_{i}\frac{\partial}{\partial\omega_{i}}\right)^{2}\Bigg]\mathcal{F}_{4}=0.
\end{equation}
\end{example}

\section{Differential Equations for MHV Gluons\label{sec:Differential-Equations-for-Gluons}}

In this section, we derive a system of partial differential equations
governing the celestial Liouville amplitude for MHV gluons. We begin
by recalling, in line with the works of \citet{stieberger2023celestial,stieberger2023yang,melton2024celestial},
how the tree-level MHV scattering amplitude for gluons in pure Yang-Mills
theory can be formulated as the semiclassical limit of Liouville vertex
operator correlation functions. Using this expression, we propose
an ansatz for the celestial Liouville amplitude of MHV gluons, corresponding
to non-zero values of the Liouville coupling constant $b$. From this
ansatz, we derive a partial differential equation via the Ward identities
of Liouville theory, suitable for perturbative solution in $b^{2}$.

The starting point of our analysis is the Parke-Taylor formula (\citet{parke1986amplitude};
see also \citet{elvang2013scattering,badger2024scattering} for modern
reviews) for the tree-level MHV scattering amplitude of $n$-gluons.
It is given by:
\begin{equation}
A_{n}=\frac{\left\langle 12\right\rangle ^{4}}{\left\langle 12\right\rangle \left\langle 23\right\rangle ...\left\langle n1\right\rangle }\delta\left(\sum^{n}_{i=1}p^{\mu}_{i}\right),\label{eq:Parke-Taylor-1}
\end{equation}
where, for simplicity, we omit the colour indices.

Substituting the spinor-helicity variables with the standard parametrisation
of null momenta in terms of the frequencies $\omega_{i}$ and the
holomorphic coordinates $z_{i}$, as employed in celestial holography,
where\footnote{We adhere to the conventions of \citet{pasterski2021lectures}.}:
\[
\left\langle ij\right\rangle \coloneqq\varepsilon_{AB}\mu^{A}_{i}\mu^{B}_{j}=\sqrt{\omega_{i}\omega_{j}}z_{ij},
\]
the Parke-Taylor amplitude becomes:
\begin{equation}
A_{n}=\frac{z^{4}_{12}}{z_{12}z_{23}...z_{n1}}\delta\left(\sum^{n}_{i=1}\omega_{i}q^{\mu}\left(z_{i},\bar{z}_{i}\right)\right)\prod^{n}_{j=1}\omega^{\alpha'_{j}}_{j},\label{eq:Parke-Taylor-2}
\end{equation}
where the exponents of the frequencies are:
\begin{equation}
\alpha'_{1}=\alpha'_{2}=1,\,\,\,\,\alpha'_{3}=...=\alpha'_{n}=-1.
\end{equation}
We use the prime notation for these exponents to distinguish them
from those appearing in the graviton amplitude.

Next, we substitute the integral representation of the energy-momentum
delta function:
\begin{equation}
\delta\left(\sum^{n}_{i=1}\omega_{i}q^{\mu}\left(z_{i},\bar{z}_{i}\right)\right)=\int\frac{d^{4}x}{\left(2\pi\right)^{4}}\prod^{n}_{i=1}e^{i\omega_{i}q\left(z_{i},\bar{z}_{i}\right)\cdot x},
\end{equation}
into the above expression, yielding:
\begin{equation}
A_{n}=\frac{z^{4}_{12}}{z_{12}z_{23}...z_{n1}}\int\frac{d^{4}x}{\left(2\pi\right)^{4}}\prod^{n}_{i=1}\omega^{\alpha'_{i}}_{i}e^{i\omega_{i}q\left(z_{i},\bar{z}_{i}\right)\cdot x}.\label{eq:Parke-Taylor-3}
\end{equation}

According to the celestial holography dictionary, the celestial amplitude
$\widehat{A}_{n}$ corresponding to the tree-level MHV $n$-gluon
amplitude is given by the $\varepsilon$-regulated Mellin transform
of Eq. (\ref{eq:Parke-Taylor-3}):
\begin{equation}
\widehat{A}_{n}\left(\left\{ \Delta_{i},z_{i},\bar{z}_{i}\right\} \right)=\prod^{n}_{i=1}\int^{\infty}_{0}d\omega_{i}\omega^{\Delta_{i}-1}_{i}e^{-\varepsilon\omega_{i}}A_{n}\left(\left\{ \omega_{i},z_{i},\bar{z}_{i}\right\} \right).
\end{equation}
Substituting the expression for $A_{n}$, we obtain:
\begin{equation}
\widehat{A}_{n}=\frac{z^{4}_{12}}{z_{12}z_{23}...z_{n1}}\int\frac{d^{4}x}{\left(2\pi\right)^{4}}\prod^{n}_{i=1}\int^{\infty}_{0}d\omega_{i}\omega^{\left(\Delta_{i}+\alpha'_{i}\right)-1}_{i}e^{-\omega_{i}\left(\varepsilon-iq\left(z_{i},\bar{z}_{i}\right)\cdot x\right)}.\label{eq:Mellin}
\end{equation}
Defining the parameters:
\begin{equation}
\rho^{(0)}_{i}\coloneqq\frac{1}{2}\left(\Delta_{i}+\alpha'_{i}\right),\qquad\rho_{i}(b)\coloneqq\rho^{(0)}_{i}+b^{2}\,\bigl((\rho^{(0)}_{i})^{2}-\rho^{(0)}_{i}\bigr)+\mathcal{O}(b^{4})\,,\label{eq:Rho}
\end{equation}
The first quantity, $\rho^{(0)}_{i}$, is the Mellin parameter in
the semiclassical approximation, while the second, $\rho_{i}(b)$,
is the finite-$b$ parameter entering the conformal dimension of the
Liouville vertex operator. (See \citet{stieberger2023yang}.) So,
the Mellin transform yields:
\begin{equation}
\widehat{A}_{n}=\frac{z^{4}_{12}}{z_{12}z_{23}...z_{n1}}\int\frac{d^{4}x}{\left(2\pi\right)^{4}}\prod^{n}_{i=1}\frac{\Gamma\left(2\rho_{i}\right)}{\left(\varepsilon-iq\left(z_{i},\bar{z}_{i}\right)\cdot x\right)^{2\rho_{i}}}.
\end{equation}

Following the same steps outlined in Eqs. (\ref{eq:Step}, \ref{eq:Step-1},
\ref{eq:Step-2}) to obtain Eq. (\ref{eq:Final}) in the gravitational
case, the spacetime integral above can be expressed as the semiclassical
limit of a Liouville field theory correlation function, as follows:
\begin{align}
 & \int\frac{d^{4}x}{\left(2\pi\right)^{4}}\prod^{n}_{i=1}\frac{\Gamma\left(2\rho_{i}\right)}{\left(\varepsilon-iq\left(z_{i},\bar{z}_{i}\right)\cdot x\right)^{2\rho_{i}}}\\
 & =\frac{1}{\left(2\pi\right)^{4}\mathcal{N}_{\mu,b}}\lim_{b\rightarrow0^{+}}\int^{\infty}_{0}d\tau\tau^{3}\left\langle \prod^{n}_{i=1}e^{-2\rho_{i}\log\tau}\Gamma\left(2\rho_{i}\right)V_{b\rho_{i}}\left(z_{i},\bar{z}_{i}\right)\right\rangle .
\end{align}
Consequently, the celestial $n$-point amplitude for tree-level MHV
scattering of gluons becomes:
\begin{equation}
\widehat{A}_{n}=\frac{1}{\left(2\pi\right)^{4}\mathcal{N}_{\mu,b}}\frac{z^{4}_{12}}{z_{12}z_{23}...z_{n1}}\lim_{b\rightarrow0^{+}}\int^{\infty}_{0}d\tau\tau^{3}\left\langle \prod^{n}_{i=1}e^{-2\rho_{i}\log\tau}\Gamma\left(2\rho_{i}\right)V_{b\rho_{i}}\left(z_{i},\bar{z}_{i}\right)\right\rangle .
\end{equation}
Using this result as an ansatz, we define the \emph{celestial Liouville
amplitude} for the scattering of MHV gluons as:
\begin{equation}
\widehat{\mathcal{A}}_{n}\coloneqq\frac{1}{\left(2\pi\right)^{4}\mathcal{N}_{\mu,b}}\frac{z^{4}_{12}}{z_{12}z_{23}...z_{n1}}\int^{\infty}_{0}d\tau\tau^{3}\left\langle \prod^{n}_{i=1}e^{-2\rho_{i}\log\tau}\Gamma\left(2\rho_{i}\right)V_{b\rho_{i}}\left(z_{i},\bar{z}_{i}\right)\right\rangle .\label{eq:Celestial}
\end{equation}

To derive a partial differential equation characterising the amplitude
$\widehat{\mathcal{A}}_{n}$, let $d'_{i}$ denote the conformal weight
of the vertex operator $\mathcal{V}_{b\rho_{i}}$ in the Liouville
sector of our theory. We adopt the prime notation to distinguish these
conformal weights from those used in the gravitational context. As
outlined in the review by \citet{harlow2011analytic}, the conformal
weight of a vertex operator $\mathcal{V}_{\alpha}$ with Liouville
momentum $\alpha$ is $\alpha\left(Q-\alpha\right)$. Therefore,
\begin{equation}
d'_{i}=\rho_{i}(b)+b^{2}\rho_{i}(b)-b^{2}\bigl(\rho_{i}(b)\bigr)^{2},
\end{equation}
and, by utilising Eq. (\ref{eq:Rho}), we find (see \citet{stieberger2023yang})\footnote{Recall that we are interested in a regime where $b$ is finite however
small in comparison to unity.}
\begin{equation}
d'_{i}=\frac{\Delta_{i}+\alpha_{i}'}{2}+\mathcal{O}(b^{4}).
\end{equation}
Hence, the sum of the conformal weights of all Liouville vertex operators
appearing in the correlation function that defines $\widehat{\mathcal{A}}_{n}$
is given by:
\begin{equation}
\sum^{n}_{i=1}d'_{i}=\frac{1}{2}\sum^{n}_{i=1}\Delta_{i}+\frac{4-n}{2}+\mathcal{O}(b^{4}).
\end{equation}

Now, on the one hand, the Ward identity for the Liouville correlation
function reads:
\begin{equation}
\sum^{n}_{i=1}z_{i}\frac{\partial}{\partial z_{i}}\left\langle \prod^{n}_{j=1}V_{b\rho_{j}}\left(z_{j},\bar{z}_{j}\right)\right\rangle =-\sum^{n}_{i=1}d'_{i}\left\langle \prod^{n}_{j=1}V_{b\rho_{j}}\left(z_{j},\bar{z}_{j}\right)\right\rangle .
\end{equation}
On the other hand, the Parke-Taylor factor $z^{4}_{12}/\left(z_{12}z_{23}...z_{n1}\right)$
multiplying the integral in Eq. (\ref{eq:Celestial}) is homogeneous
in the holomorphic variables with degree $4-n$. Thus, Euler theorem
for homogeneous functions implies:
\begin{equation}
\sum^{n}_{i=1}z_{i}\frac{\partial}{\partial z_{i}}\left(\frac{z^{4}_{12}}{z_{12}z_{23}...z_{n1}}\right)=\left(4-n\right)\frac{z^{4}_{12}}{z_{12}z_{23}...z_{n1}}.
\end{equation}
Therefore, applying the differential operator $\sum_{i}z_{i}\partial/\partial z_{i}$
to Eq. (\ref{eq:Celestial}) results in the following differential
equation in Mellin space:
\begin{equation}
\left(\sum^{n}_{i=1}z_{i}\frac{\partial}{\partial z_{i}}+n-4\right)\widehat{\mathcal{A}}_{n}+\sum^{n}_{i=1}d'_{i}\widehat{\mathcal{A}}_{n}=0.\label{eq:PDE-Mellin-Space}
\end{equation}

Next, to return to momentum space, we must apply the inverse Mellin
transform to the above equation. But, if $\hat{f}\left(\Delta_{1},...,\Delta_{n}\right)$
is the Mellin transform of $f\left(\omega_{1},...,\omega_{n}\right)$,
there exists a differential operator $\mathcal{D}'_{\left(n\right)}$,
linear with respect to the frequencies, such that:
\begin{equation}
\prod^{n}_{i=1}\int^{c+i\infty}_{c-i\infty}\frac{d\Delta_{i}}{2\pi i}\omega^{\Delta_{i}-1}_{i}\sum^{n}_{j=1}d'_{j}\hat{f}\left(\Delta_{1},...,\Delta_{n}\right)=\mathcal{D}'_{\left(n\right)}f\left(\omega_{1},...,\omega_{n}\right),
\end{equation}
where:
\begin{equation}
\mathcal{D}'_{\left(n\right)}\coloneqq\frac{4-n}{2}-\frac{1}{2}\sum^{n}_{i=1}\omega_{i}\frac{\partial}{\partial\omega_{i}}+\mathcal{O}(b^{4})\label{eq:Operator}
\end{equation}
As we have done throughout this section, the prime in the above operator
serves to distinguish it from the differential operator associated
with the system of equations governing graviton amplitudes.

Thus, the inverse Mellin transform of Eq. (\ref{eq:PDE-Mellin-Space})
yields the partial differential equation that characterises the celestial
Liouville amplitude for MHV gluons:
\begin{equation}
\left(\sum^{n}_{i=1}z_{i}\frac{\partial}{\partial z_{i}}+n-4\right)\mathcal{A}_{n}+\mathcal{D}'_{\left(n\right)}\mathcal{A}_{n}=0.
\end{equation}
Employing the decomposition of the differential operator $\mathcal{D}'_{\left(n\right)}$,
as defined in Eq. (\ref{eq:Operator}), we can fully express our equation
as:
\begin{equation}
\left[\sum^{n}_{i=1}z_{i}\frac{\partial}{\partial z_{i}}-\frac{1}{2}\sum^{n}_{i=1}\omega_{i}\frac{\partial}{\partial\omega_{i}}+\frac{n-4}{2}\right]\mathcal{A}_{n}=\mathcal{O}(b^{4})\,.\label{eq:PDE-3}
\end{equation}

To solve this equation perturbatively in $b^{2}$, it is useful to
consider the following regime. We assume that $0<b^{2}\ll1$ and retain
terms through the first non-trivial order in $b^{2}$. In this limit,
we discard terms of order $b^{4}$ and higher, so that Eq. (\ref{eq:PDE-3})
simplifies to:
\begin{equation}
\sum^{n}_{i=1}\left(z_{i}\frac{\partial}{\partial z_{i}}-\frac{1}{2}\omega_{i}\frac{\partial}{\partial\omega_{i}}\right)\mathcal{A}_{n}=\frac{4-n}{2}\mathcal{A}_{n}+\mathcal{O}(b^{4}).\label{eq:PDE-4}
\end{equation}

\section{OPE Structure of Celestial Liouville Amplitudes for Gravitons\label{sec:Celestial-OPE-Gravitons}}

In this section, we solve the system of partial differential equations
in Eq. (\ref{eq:System}) perturbatively, up to leading order in $b^{2}$.
We find that the corrections to the MHV $n$-graviton amplitudes exhibit
a logarithmic dependence. Subsequently, we utilise the renormalised
amplitude to compute the leading-order correction to the celestial
OPE, employing the collinear limit approach within the splitting function
formalism. The resulting celestial OPE is shown to be isomorphic to
that obtained by incorporating the one-loop correction to Einstein
gravity.

\subsection{Perturbative Expansion of Celestial Liouville Amplitudes in $b^{2}$}

To obtain a perturbative solution for the celestial Liouville amplitude
$\mathcal{M}_{n}$ for the scattering of MHV $n$-gravitons, parametrised
by the Liouville coupling constant $b$, we employ the following approach.
The amplitude $\mathcal{M}_{n}$ has a generating function $\mathcal{F}_{n}$
defined by Eqs. (\ref{eq:Generating-Function}, \ref{eq:Generating-Function-1}),
which satisfies Eq. (\ref{eq:PDE-1}) found in Section \ref{sec:Gravitons}:
\begin{equation}
\sum^{n}_{i=1}\left(z_{i}\frac{\partial}{\partial z_{i}}+n-8\right)\mathcal{F}_{n}+\mathcal{D}_{\left(n\right)}\mathcal{F}_{n}=0.\label{eq:PDE}
\end{equation}
The differential operator $\mathcal{D}_{\left(n\right)}$, defined
in Eq. (\ref{eq:Differential-Operator}), is given by:
\begin{equation}
\mathcal{D}_{\left(n\right)}\coloneqq c_{n}-\frac{1}{2}\sum^{n}_{i=1}\beta^{\left(n\right)}_{i}\omega_{i}\frac{\partial}{\partial\omega_{i}}-\frac{b^{2}}{4}\sum^{n}_{i=1}\left(\omega_{i}\frac{\partial}{\partial\omega_{i}}\right)^{2}.
\end{equation}
The constants $\beta^{\left(n\right)}_{i}$ are fully determined by
Eqs. (\ref{eq:Sigma}, \ref{eq:Sigma-1}, \ref{eq:Sigma-2}). Once
Eq. (\ref{eq:PDE}) is solved, we apply the inverse Mellin transform
to Eq. (\ref{eq:Celestial-Liouville-Amplitude-1}) to obtain $\mathcal{M}_{n}$
as:

\begin{equation}
\mathcal{M}_{n}=\widetilde{\mathscr{L}}^{\left(n\right)}_{\omega_{1},...,\omega_{n}}\mathcal{F}_{n},\label{eq:Convolution}
\end{equation}
where $\widetilde{\mathscr{L}}^{\left(n\right)}_{\omega_{1},...,\omega_{n}}$
is the operator linear in the frequencies $\omega_{1},...,\omega_{n}$,
as defined in Eq. (\ref{eq:Inverse-Mellin-Transformed-Operator}).

To determine the leading $\mathcal{O}(b^{2})$ correction to $\mathcal{M}_{n}$,
which is necessary for evaluating the correction to the celestial
OPE due to a non-zero Liouville coupling constant, we examine the
following limit. Let there exist a real number $\varepsilon>0$ and
a positive integer $N$ such that $\varepsilon<b^{2}<e^{-N}$. This
choice ensures that $b^{2}$ remains positive and sufficiently small
relative to unity, as $\varepsilon$ approaches zero and $N$ becomes
large. Thus, according to Eqs. (\ref{eq:Sigma}, \ref{eq:Sigma-1},
\ref{eq:Sigma-2}), we have $\beta^{\left(n\right)}_{i}=1+\mathcal{O}(b^{2})$
for all $1\leq i\leq n$. In this limit, Eq. (\ref{eq:PDE}) simplifies
to:
\begin{equation}
\sum^{n}_{i=1}\left(z_{i}\frac{\partial}{\partial z_{i}}-\frac{1}{2}\omega_{i}\frac{\partial}{\partial\omega_{i}}\right)\mathcal{F}_{n}=\gamma_{n}\mathcal{F}_{n}+\frac{b^{2}}{4}\sum^{n}_{i=1}\left(\omega_{i}\frac{\partial}{\partial\omega_{i}}\right)^{2}\mathcal{F}_{n}.
\end{equation}
where we defined $\gamma_{n}\coloneqq8-n-c_{n}$ to simplify our notation. 

To streamline our calculations, we assume, while this is not essential
from a physical perspective, that the number $n$ of scattered MHV
gravitons is large. This assumption is warranted as follows: Recall
that the parameter $\sigma_{i}$, as defined in Eqs. (\ref{eq:Sigma},
\ref{eq:Sigma-1}), is related to the conformal weight $\Delta_{i}$
by $2\sigma_{i}=\Delta_{i}+\alpha_{i}$ for $2\leq i\leq n-2$ and
$2\sigma_{i}=\Delta_{i}+\alpha_{i}-1$ for $i=1,n-1,n-2$, where $\alpha_{i}$
denotes the exponent of the frequency $\omega_{i}$ in the definitions
of the Mellin-transformed operators $\widehat{\mathcal{Q}}$ and $\widehat{\mathcal{P}}$
(see Eqs. (\ref{eq:Mellin-Transformed-Q-Operator}, \ref{eq:Mellin-Transformed-P-Operator})),
with $\alpha_{1}=\alpha_{2}=3$ and $\alpha_{k}=-1$ for all $k\geq3$.
Thus, for large $n$, $\sum^{n}_{i=1}\alpha^{2}_{i}\sim n$. Given
that our primary aim is to examine the collinear limit of the celestial
Liouville amplitude and to determine the celestial OPE using the splitting
function formalism, the assumption of a large number of scattered
gravitons does not diminish the generality of our results. 

Now, consider the perturbative expansion of the generating function
$\mathcal{F}_{n}$ in powers of the squared Liouville coupling constant
$b^{2}$:
\begin{equation}
\mathcal{F}_{n}=\sum^{\infty}_{k=0}b^{2k}\mathcal{F}^{k}_{n}=\mathcal{F}^{\left(0\right)}_{n}+b^{2}\mathcal{F}^{\left(1\right)}_{n}+\mathcal{O}\left(b^{4}\right).
\end{equation}
The leading term $\mathcal{F}^{\left(0\right)}_{n}$ in this expansion
satisfies the semiclassical limit $b\rightarrow0^{+}$ of Liouville
theory, which dictates that:
\begin{equation}
\sum^{n}_{i=1}\left(z_{i}\frac{\partial}{\partial z_{i}}-\frac{1}{2}\omega_{i}\frac{\partial}{\partial\omega_{i}}\right)\mathcal{F}^{\left(0\right)}_{n}=\gamma_{n}\mathcal{F}^{\left(0\right)}_{n}.\label{eq:Step-5}
\end{equation}
Then, the leading $\mathcal{O}(b^{2})$ correction satisfies:
\begin{equation}
\sum^{n}_{i=1}\left(z_{i}\frac{\partial}{\partial z_{i}}-\frac{1}{2}\omega_{i}\frac{\partial}{\partial\omega_{i}}\right)\mathcal{F}^{\left(1\right)}_{n}=\gamma_{n}\mathcal{F}^{\left(1\right)}_{n}+\frac{b^{2}}{4}\sum^{n}_{i=1}\left(\omega_{i}\frac{\partial}{\partial\omega_{i}}\right)^{2}\mathcal{F}^{\left(0\right)}_{n}.\label{eq:Step-3}
\end{equation}

To solve Eq. (\ref{eq:Step-3}), we introduce the ansatz:
\begin{equation}
\mathcal{F}^{\left(1\right)}_{n}\left(\omega,z\right)=\mathcal{F}^{\left(0\right)}_{n}\left(\omega,z\right)G_{n}\left(\omega,z\right).\label{eq:Step-4}
\end{equation}
Substituting this ansatz into Eq. (\ref{eq:Step-3}), and applying
the condition specified by Eq. (\ref{eq:Step-5}) for the leading
term, we derive:
\begin{equation}
\mathcal{F}^{\left(0\right)}_{n}\sum^{n}_{i=1}\left(z_{i}\frac{\partial}{\partial z_{i}}-\frac{1}{2}\omega_{i}\frac{\partial}{\partial\omega_{i}}\right)G_{n}=\frac{b^{2}}{4}\sum^{n}_{i=1}\left(\omega_{i}\frac{\partial}{\partial\omega_{i}}\right)^{2}\mathcal{F}^{\left(0\right)}_{n}.\label{eq:Step-6}
\end{equation}
Since the leading-order term $\mathcal{F}^{\left(0\right)}_{n}$ arises
from the semiclassical limit of the Liouville sector of our theory,
it follows from Eqs. (\ref{eq:Celestial-Liouville-Amplitude}, \ref{eq:Celestial-Liouville-Amplitude-1})
that:
\begin{align}
\mathcal{F}^{\left(0\right)}_{n} & =\frac{1}{\left(2\pi\right)^{4}\mathcal{N}_{\mu,b}}\frac{z^{8}_{12}}{z_{12}z_{23}...z_{n1}}\left(\prod^{n}_{i=1}\int^{c+i\infty}_{c-i\infty}\frac{d\Delta_{i}}{2\pi i}\omega^{-\Delta_{i}}_{i}\right)\\
 & \lim_{b\rightarrow0^{+}}\int^{\infty}_{0}d\tau\tau^{3}\left\langle \prod^{n}_{j=1}e^{-2\sigma_{j}\log\tau}\Gamma\left(2\sigma_{j}\right)V_{b\sigma_{j}}\left(z_{j},\bar{z}_{j}\right)\right\rangle +\mathcal{P}\left(2,...,n-2\right)+\left(\left(\bar{z}\rightarrow-\bar{z}\right)\right).
\end{align}
By reversing the steps leading from Eqs. (\ref{eq:Step}, \ref{eq:Step-1},
\ref{eq:Step-2}) to Eq. (\ref{eq:Final}), we obtain:
\begin{equation}
\mathcal{F}^{\left(0\right)}_{n}=\frac{z^{8}_{12}}{z_{12}z_{23}...z_{n1}}\left(\prod^{n}_{i=1}\int^{c+i\infty}_{c-i\infty}\frac{d\Delta_{i}}{2\pi i}\omega^{-\Delta_{i}}_{i}\right)\int\frac{d^{4}x}{\left(2\pi\right)^{4}}\prod^{n}_{j=1}\frac{\Gamma\left(2\sigma_{j}\right)}{\left(\varepsilon-iq\left(z_{j},\bar{z}_{j}\right)\cdot x\right)^{2\sigma_{j}}}.\label{eq:Step-7}
\end{equation}
Substituting the identity:
\begin{equation}
\prod^{n}_{i=1}\frac{\Gamma\left(2\sigma_{i}\right)}{\left(\varepsilon-iq\left(z_{i},\bar{z}_{i}\right)\cdot x\right)^{2\sigma_{i}}}=\left(\prod^{n}_{i=1}\int^{\infty}_{0}d\omega_{i}\omega^{2\sigma_{i}-1}_{i}\right)\prod^{n}_{j=1}e^{-\omega_{j}\left(\varepsilon-iq\left(z_{j},\bar{z}_{j}\right)\cdot x\right)},
\end{equation}
into Eq. (\ref{eq:Step-7}), we find:
\begin{equation}
\mathcal{F}^{\left(0\right)}_{n}=\frac{z^{8}_{12}}{z_{12}z_{23}...z_{n1}}\prod^{n}_{i=1}e^{\alpha_{i}\log\omega_{i}}\int\frac{d^{4}x}{\left(2\pi\right)^{4}}e^{i\sum^{n}_{k=1}\omega_{k}q\left(z_{k},\bar{z}_{k}\right)\cdot x}.
\end{equation}
Thus, under the assumption of large $n$, we have:
\begin{equation}
\sum^{n}_{i=1}\left(\omega_{i}\frac{\partial}{\partial\omega_{i}}\right)^{2}\mathcal{F}^{\left(0\right)}_{n}=\sum^{n}_{i=1}\alpha^{2}_{i}\mathcal{F}^{\left(0\right)}_{n}\sim n\mathcal{F}^{\left(0\right)}_{n}.
\end{equation}
Therefore, Eq. (\ref{eq:Step-6}) simplifies to:
\begin{equation}
\sum^{n}_{i=1}\left(z_{i}\frac{\partial}{\partial z_{i}}-\frac{1}{2}\omega_{i}\frac{\partial}{\partial\omega_{i}}\right)G_{n}=\frac{nb^{2}}{4}.
\end{equation}
The solution that is consistent with global conformal transformations
and Lorentz invariance is given by:
\[
G_{n}=\frac{b^{2}}{4}\sum_{j=i+1}\log\frac{\omega_{i}\omega_{j}\left|z_{ij}\right|^{2}}{m^{2}},
\]
where $m^{2}$ is a renormalisation parameter with units of energy
squared, and we have set $z_{n+1}\coloneqq z_{1}$ to simplify the
summation. 

Consequently, the generating function, up to the leading $\mathcal{O}(b^{2})$
correction, takes the following form:
\begin{equation}
\mathcal{F}_{n}=\left(1-\frac{b^{2}}{4}\sum_{j=i+1}\log\frac{\omega_{i}\omega_{j}\left|z_{ij}\right|^{2}}{m^{2}}\right)\mathcal{F}^{\left(0\right)}_{n}+\mathcal{O}(b^{4}),
\end{equation}
and Eq. (\ref{eq:Convolution}) ultimately furnishes:
\begin{equation}
\mathcal{M}_{n}=\left(1-\frac{b^{2}}{4}\sum_{j=i+1}\log\frac{\omega_{i}\omega_{j}\left|z_{ij}\right|^{2}}{m^{2}}\right)\mathcal{M}^{\left(0\right)}_{n}+\mathcal{O}(b^{4}).
\end{equation}

Finally, noting that $\mathcal{M}^{\left(0\right)}_{n}$ corresponds
to the semiclassical approximation of the Liouville sector of our
theory, it is identified with the tree-level MHV $n$-graviton amplitude,
$\mathcal{M}^{\left(0\right)}_{n}=M_{n}$. Therefore, the celestial
Liouville amplitude for MHV graviton scattering may be succinctly
represented as:
\begin{equation}
\mathcal{M}_{n}=\mathcal{R}_{n}M_{n}+\mathcal{O}(b^{2}),\label{eq:Renormalised-Amplitude}
\end{equation}
where the renormalisation factor is given by:
\begin{equation}
\mathcal{R}_{n}\coloneqq1-\frac{b^{2}}{4}\sum_{1\leq i<j\leq n}\log\frac{\omega_{i}\omega_{j}\left|z_{ij}\right|^{2}}{m^{2}}.\label{eq:Renormalisation-Factor}
\end{equation}

\subsection{Correction to Celestial OPE for Gravitons via Renormalised Liouville
Amplitudes}

In this subsection, we employ the renormalised celestial Liouville
amplitude derived above to determine the correction to the celestial
OPE for gravitons, parametrised by the Liouville coupling constant
$b$. This is accomplished through the splitting function formalism,
which is used to examine the collinear limit of scattering amplitudes,
as developed by \citet{bern1999multi,white2011factorization,akhoury2011collinear}
(for a pedagogical review, see \citet{elvang2013scattering,badger2024scattering}). 

Celestial OPEs obtained from soft and collinear limits were introduced
in \citet{fan2019soft} and subsequently developed in the extended
BMS and supersymmetric BMS analyses of \citet{fotopoulos2020extended}.
We shall use the splitting-function presentation reviewed and systematised
in \citet{pate2021celestial}.

Within the framework of celestial holography, the method of computing
the celestial OPE via the collinear limit and the splitting function
formalism was first introduced by \citet{pate2021celestial}.

According to this formalism, the holomorphic collinear limit of the
$n$-graviton amplitude $M_{n}$ as $z_{12}\rightarrow0$ factorises
in the following manner:
\begin{align}
 & \lim_{z_{12}\rightarrow0}M_{n}\left(\omega_{1},z_{1};\omega_{2},z_{2};...;\omega_{n},z_{n}\right)\\
 & =\sum_{s=\pm2}\text{Split}^{s}_{s_{1}s_{2}}\left(p_{1},p_{2}\right)M_{n-1}\left(\omega_{1}+\omega_{2},z_{2};\omega_{3},z_{3};...;\omega_{n},z_{n}\right).\label{eq:Collinear-Limit}
\end{align}
where the universal collinear factor, $\text{Split}^{s}_{s_{1}s_{2}}\left(p_{1},p_{2}\right)$,
is referred to as the ``splitting function.'' The only non-vanishing
combination of helicities for $\text{Split}^{s}_{s_{1},s_{2}}\left(p_{1},p_{2}\right)$
are:
\begin{equation}
\text{Split}^{2}_{2,2}\left(p_{1},p_{2}\right)=-\frac{\kappa}{2}\frac{\bar{z}_{12}}{z_{12}}\frac{\left(\omega_{1}+\omega_{2}\right)^{2}}{\omega_{1}\omega_{2}},\label{eq:Split}
\end{equation}
and:
\begin{equation}
\text{Split}^{-2}_{2,-2}\left(p_{1},p_{2}\right)=-\frac{\kappa}{2}\frac{\bar{z}_{12}}{z_{12}}\frac{\omega^{3}_{2}}{\omega_{1}\left(\omega_{1}+\omega_{2}\right)^{2}}.\label{eq:Split-1}
\end{equation}

In what follows, we will focus exclusively on the holomorphic collinear
limit. Thus, in Eq. (\ref{eq:Collinear-Limit}), it is to be understood
that $\bar{z}_{12}$ is held fixed as $z_{12}\rightarrow0$. Moreover,
for simplicity, sometimes we may omit the anti-holomorphic variables
from the arguments, while retaining explicit dependence on the frequencies
$\omega_{1},...,\omega_{n}$ and holomorphic variables $z_{1},...,z_{n}$
to carefully track their transformations in the collinear limit.

We begin by considering the collinear limit of two positive helicity
gravitons, where $s_{1}=s_{2}=2$. From Eqs. (\ref{eq:Collinear-Limit},
\ref{eq:Split}), we have:
\begin{align}
 & \lim_{z_{12}\rightarrow0}M_{n}\left(\omega_{1},z_{1};\omega_{2},z_{2};...;\omega_{n},z_{n}\right)\\
 & =-\frac{\kappa}{2}\frac{\bar{z}_{12}}{z_{12}}\frac{\left(\omega_{1}+\omega_{2}\right)^{2}}{\omega_{1}\omega_{2}}M_{n-1}\left(\omega_{1}+\omega_{2},z_{2};\omega_{3},z_{3};...;\omega_{n},z_{n}\right).\label{eq:Collinear-Limit-1}
\end{align}

On the other hand, the renormalisation factor, as defined in Eq. (\ref{eq:Renormalisation-Factor}),
may be expressed as:
\begin{align}
 & \mathcal{R}_{n}\left(\omega_{1},z_{1};\omega_{2},z_{2};...;\omega_{n},z_{n}\right)\\
 & =\mathcal{R}_{n-1}\left(\omega_{1}+\omega_{2},z_{2};\omega_{3},z_{3};...;\omega_{n},z_{n}\right)+\frac{b^{2}}{2}\log\frac{\omega_{1}\omega_{2}\left|z_{12}\right|}{\left(\omega_{1}+\omega_{2}\right)m}.\label{eq:Renormalisation-Factor-1}
\end{align}

Combining Eqs. (\ref{eq:Renormalised-Amplitude}, \ref{eq:Renormalisation-Factor-1}),
we obtain the collinear limit of the renormalised celestial Liouville
amplitude as follows:
\begin{align}
 & \lim_{z_{12}\rightarrow0}\mathcal{M}_{n}\left(\omega_{1},z_{1};\omega_{2},z_{2};...;\omega_{n},z_{n}\right)\\
 & =\mathcal{R}_{n-1}\left(\omega_{1}+\omega_{2},z_{2};\omega_{3},z_{3};...;\omega_{n},z_{n}\right)\lim_{z_{12}\rightarrow0}M_{n}\left(\omega_{1},z_{1};\omega_{2},z_{2};...;\omega_{n},z_{n}\right)\\
 & +\frac{b^{2}}{2}\lim_{z_{12}\rightarrow0}\left(M_{n}\left(\omega_{1},z_{1};\omega_{2},z_{2};...;\omega_{n},z_{n}\right)\log\frac{\omega_{1}\omega_{2}\left|z_{12}\right|}{\left(\omega_{1}+\omega_{2}\right)m}\right)+\mathcal{O}\left(b^{2}\right)
\end{align}
Substituting Eq. (\ref{eq:Collinear-Limit-1}), we find:
\begin{align}
 & \lim_{z_{12}\rightarrow0}\mathcal{M}_{n}\left(\omega_{1},z_{1};\omega_{2},z_{2};...;\omega_{n},z_{n}\right)\\
 & =-\frac{\kappa}{2}\frac{\bar{z}_{12}}{z_{12}}\frac{\left(\omega_{1}+\omega_{2}\right)^{2}}{\omega_{1}\omega_{2}}\mathcal{R}_{n-1}\left(\omega_{1}+\omega_{2},z_{2};\omega_{3},z_{3};...;\omega_{n},z_{n}\right)M_{n-1}\left(\omega_{1}+\omega_{2},z_{2};...;\omega_{n},z_{n}\right)\\
 & -\frac{\kappa b^{2}}{4}\frac{\bar{z}_{12}}{z_{12}}\frac{\left(\omega_{1}+\omega_{2}\right)^{2}}{\omega_{1}\omega_{2}}M_{n-1}\left(\omega_{1}+\omega_{2},z_{2};\omega_{3},z_{3};...;\omega_{n},z_{n}\right)\log\frac{\omega_{1}\omega_{2}\left|z_{12}\right|}{\left(\omega_{1}+\omega_{2}\right)m}+\mathcal{O}\left(b^{2}\right).
\end{align}
Finally, as we interested only in the terms up to $\mathcal{O}(b^{2})$,
we have:
\begin{align}
 & \lim_{z_{12}\rightarrow0}\mathcal{M}_{n}\left(\omega_{1},z_{1};\omega_{2},z_{2};...;\omega_{n},z_{n}\right)\\
 & =-\frac{\kappa}{2}\frac{\bar{z}_{12}}{z_{12}}\frac{\left(\omega_{1}+\omega_{2}\right)^{2}}{\omega_{1}\omega_{2}}\mathcal{M}_{n-1}\left(\omega_{1}+\omega_{2},z_{2};...;\omega_{n},z_{n}\right)\\
 & -\frac{\kappa b^{2}}{4}\frac{\bar{z}_{12}}{z_{12}}\frac{\left(\omega_{1}+\omega_{2}\right)^{2}}{\omega_{1}\omega_{2}}\mathcal{M}_{n-1}\left(\omega_{1}+\omega_{2},z_{2};\omega_{3},z_{3};...;\omega_{n},z_{n}\right)\log\frac{\omega_{1}\omega_{2}\left|z_{12}\right|}{\left(\omega_{1}+\omega_{2}\right)m}+\mathcal{O}\left(b^{2}\right).
\end{align}
The collinear limit of the celestial amplitude is obtained by performing
the Mellin transform of the above expression, yielding:
\begin{align}
 & \lim_{z_{12}\rightarrow0}\widehat{\mathcal{M}}_{n}\left(\Delta_{1},z_{1};\Delta_{2},z_{2};...;\Delta_{n},z_{n}\right)\\
 & =-\frac{\kappa}{2}\frac{\bar{z}_{12}}{z_{12}}\left(\prod^{n}_{i=1}\int^{\infty}_{0}d\omega_{i}\omega^{\Delta_{i}-1}_{i}\right)\frac{\left(\omega_{1}+\omega_{2}\right)^{2}}{\omega_{1}\omega_{2}}\mathcal{M}_{n-1}\left(\omega_{1}+\omega_{2},z_{2};...;\omega_{n},z_{n}\right)\\
 & -\frac{\kappa b^{2}}{4}\frac{\bar{z}_{12}}{z_{12}}\left(\prod^{n}_{i=1}\int^{\infty}_{0}d\omega_{i}\omega^{\Delta_{i}-1}_{i}\right)\frac{\left(\omega_{1}+\omega_{2}\right)^{2}}{\omega_{1}\omega_{2}}\log\frac{\omega_{1}\omega_{2}\left|z_{12}\right|}{\left(\omega_{1}+\omega_{2}\right)m}\\
 & \mathcal{M}_{n-1}\left(\omega_{1}+\omega_{2},z_{2};\omega_{3},z_{3};...;\omega_{n},z_{n}\right)+\mathcal{O}\left(b^{2}\right).
\end{align}

To evaluate these Mellin integrals, we introduce the change of variables
$\omega_{1}=t\omega_{P}$ and $\omega_{2}=\left(1-t\right)\omega_{P}$,
with $t\in\left[0,1\right]$ and $\omega_{P}\in\text{[0,\ensuremath{\infty})}.$
This gives:
\begin{equation}
d\omega_{1}d\omega_{2}\omega^{\Delta_{1}-1}_{1}\omega^{\Delta_{2}-1}_{2}\frac{\left(\omega_{1}+\omega_{2}\right)^{2}}{\omega_{1}\omega_{2}}=dtd\omega_{p}t^{\left(\Delta_{1}-1\right)-1}\left(1-t\right)^{\left(\Delta_{2}-1\right)-1}\omega^{\left(\Delta_{1}+\Delta_{2}\right)-1}_{P}.
\end{equation}
Using the identity for the partial derivative with respect to the
conformal weight of the Mellin transform:
\begin{equation}
\int^{\infty}_{0}d\omega\omega^{\Delta-1}f\left(\omega\right)\log\omega=\frac{\partial}{\partial\Delta}\int^{\infty}_{0}d\omega\omega^{\Delta-1}f\left(\omega\right),
\end{equation}
we conclude:
\begin{align}
 & \lim_{z_{12}\rightarrow0}\widehat{\mathcal{M}}_{n}\left(\Delta_{1},z_{1};\Delta_{2},z_{2};...;\Delta_{n},z_{n}\right)\\
 & =-\frac{\kappa}{2}\frac{\bar{z}_{12}}{z_{12}}B\left(\Delta_{1}-1,\Delta_{2}-1\right)\widehat{\mathcal{M}}_{n-1}\left(\Delta_{1}+\Delta_{2},z_{2};\Delta_{3},z_{3};...;\Delta_{n},z_{n}\right)-\frac{\kappa b^{2}}{4}\frac{\bar{z}_{12}}{z_{12}}\\
 & \left(\frac{\partial}{\partial\Delta_{1}}+\frac{\partial}{\partial\Delta_{2}}+\frac{1}{2}\log\frac{\left|z_{12}\right|^{2}}{m^{2}}\right)B\left(\Delta_{1}-1,\Delta_{2}-1\right)\widehat{\mathcal{M}}_{n-1}\left(\Delta_{1}+\Delta_{2},z_{2};...;\Delta_{n},z_{n}\right)\\
 & +\frac{\kappa b^{2}}{4}\frac{\bar{z}_{12}}{z_{12}}B\left(\Delta_{1}-1,\Delta_{2}-1\right)\frac{\partial}{\partial\Delta}\bigg|_{\Delta=\Delta_{1}+\Delta_{2}}\widehat{\mathcal{M}}_{n-1}\left(\Delta_{1}+\Delta_{2},z_{2};\Delta_{3},z_{3};...;\Delta_{n},z_{n}\right).
\end{align}

Finally, let $G^{+}_{\Delta}\left(z,\bar{z}\right)$ denote the conformal
primary corresponding to positive helicity gravitons in the celestial
CFT. We conclude that the leading-order correction in $b^{2}$ to
the celestial OPE is:
\begin{align}
 & G^{+}_{\Delta_{1}}\left(z_{1},\bar{z}_{1}\right)G^{+}_{\Delta_{2}}\left(z_{2},\bar{z}_{2}\right)\sim\\
 & -\frac{\kappa}{2}\frac{\bar{z}_{12}}{z_{12}}B\left(\Delta_{1}-1,\Delta_{2}-1\right)G^{+}_{\Delta_{1}+\Delta_{2}}\left(z_{2},\bar{z}_{2}\right)\\
 & -\frac{\kappa b^{2}}{4}\left(\frac{\partial}{\partial\Delta_{1}}+\frac{\partial}{\partial\Delta_{2}}+\frac{1}{2}\log\frac{\left|z_{12}\right|^{2}}{m^{2}}\right)B\left(\Delta_{1}-1,\Delta_{2}-1\right)G^{+}_{\Delta_{1}+\Delta_{2}}\left(z_{2},\bar{z}_{2}\right)\\
 & +\frac{\kappa b^{2}}{4}\frac{\bar{z}_{12}}{z_{12}}B\left(\Delta_{1}-1,\Delta_{2}-1\right)\frac{\partial}{\partial\Delta}\bigg|_{\Delta=\Delta_{1}+\Delta_{2}}G^{+}_{\Delta}\left(z_{2},\bar{z}_{2}\right).
\end{align}

Applying the same method to the OPE between a positive helicity graviton
$G^{+}_{\Delta_{1}}\left(z_{1},\bar{z}_{1}\right)$ and a negative
helicity graviton $G^{-}_{\Delta_{2}}\left(z_{2},\bar{z}_{2}\right)$,
we derive:
\begin{align}
 & G^{+}_{\Delta_{1}}\left(z_{1},\bar{z}_{1}\right)G^{-}_{\Delta_{2}}\left(z_{2},\bar{z}_{2}\right)\sim\\
 & -\frac{\kappa}{2}\frac{\bar{z}_{12}}{z_{12}}B\left(\Delta_{1}-1,\Delta_{2}+3\right)G^{-}_{\Delta_{1}+\Delta_{2}}\left(z_{2},\bar{z}_{2}\right)\\
 & -\frac{\kappa b^{2}}{4}\frac{\bar{z}_{12}}{z_{12}}\left(\frac{\partial}{\partial\Delta_{1}}+\frac{\partial}{\partial\Delta_{2}}+\frac{1}{2}\log\frac{\left|z_{12}\right|^{2}}{m^{2}}\right)B\left(\Delta_{1}-1,\Delta_{2}+3\right)G^{-}_{\Delta_{1}+\Delta_{2}}\left(z_{2},\bar{z}_{2}\right)\\
 & +\frac{\kappa b^{2}}{4}\frac{\bar{z}_{12}}{z_{12}}B\left(\Delta_{1}-1,\Delta_{2}+3\right)\frac{\partial}{\partial\Delta}\bigg|_{\Delta=\Delta_{1}+\Delta_{2}}G^{-}_{\Delta}\left(z_{2},\bar{z}_{2}\right).
\end{align}

Consequently, the corrections to the celestial OPE induced by a non-zero
Liouville coupling constant are isomorphic to those arising from the
celestial OPE when accounting for the one-loop correction to Einstein
gravity, as studied by \citet{bhardwaj2022loop,bhardwaj2024celestial,bittleston2023associativity,krishna2024celestial}.

\section{OPE Structure of Celestial Liouville Amplitudes for Gluons\label{sec:Celestial-OPE-Gluons}}

In this section, we examine the perturbative solution of the partial
differential equation obtained in Eq. (\ref{eq:System}) of Section
\ref{sec:Differential-Equations-for-Gluons}, considering terms up
to leading-order in $b^{2}$. Similar to the gravitational case discussed
in the previous section, the corrections to the MHV $n$-gluon amplitudes
exhibit a logarithmic dependence. Using the renormalised amplitude,
we then compute the leading-order correction to the celestial OPE,
employing the collinear limit approach. The resulting celestial OPE
is shown to be isomorphic to that obtained by incorporating the one-loop
correction in pure Yang-Mills theory as studied by \citet{bhardwaj2022loop,banerjee2024all,adamo2023all,krishna2024celestial,bittleston2023associativity}.

\subsection{Perturbative Expansion of Celestial Liouville Amplitudes in $b^{2}$}

Here, we consider perturbative solutions to the equation characterising
the celestial Liouville amplitude for MHV gluons, in the limit discussed
at the end of Section \ref{sec:Differential-Equations-for-Gluons}.:
\begin{equation}
\sum^{n}_{i=1}\left(z_{i}\frac{\partial}{\partial z_{i}}-\frac{1}{2}\omega_{i}\frac{\partial}{\partial\omega_{i}}\right)\mathcal{A}_{n}=\frac{4-n}{2}\mathcal{A}_{n}+\mathcal{O}(b^{4}).
\end{equation}
We seek solutions as a power series expansion in the Liouville coupling
constant:
\begin{equation}
\mathcal{A}_{n}=\sum^{\infty}_{k=0}b^{2k}\mathcal{A}^{\left(k\right)}_{n}=\mathcal{A}^{\left(0\right)}_{n}+b^{2}\mathcal{A}^{\left(1\right)}_{n}+\mathcal{O}(b^{4}),
\end{equation}
where the leading term obeys the semiclassical equation:
\begin{equation}
\sum^{n}_{i=1}\left(z_{i}\frac{\partial}{\partial z_{i}}-\frac{1}{2}\omega_{i}\frac{\partial}{\partial\omega_{i}}\right)\mathcal{A}^{\left(0\right)}_{n}=\frac{n-4}{2}\mathcal{A}^{\left(0\right)}_{n}.
\end{equation}
The $\mathcal{O}(b^{2})$ term satisfies:
\begin{equation}
\sum^{n}_{i=1}\left(z_{i}\frac{\partial}{\partial z_{i}}-\frac{1}{2}\omega_{i}\frac{\partial}{\partial\omega_{i}}\right)\mathcal{A}^{\left(1\right)}_{n}=\frac{n-4}{2}\mathcal{A}^{\left(1\right)}_{n}+\frac{b^{2}}{4}\sum^{n}_{i=1}\left(\omega_{i}\frac{\partial}{\partial\omega_{i}}\right)^{2}\mathcal{A}^{\left(0\right)}_{n}.\label{eq:PDE-5}
\end{equation}

As in the gravitational case, we adopt the ansatz:
\begin{equation}
\mathcal{A}^{\left(1\right)}_{n}\left(\left\{ \omega_{i},z_{i}\right\} \right)=\mathcal{A}^{\left(0\right)}_{n}\left(\left\{ \omega_{i},z_{i}\right\} \right)H_{n}\left(\left\{ \omega_{i},z_{i}\right\} \right).
\end{equation}
Thus, Eq. (\ref{eq:PDE-5}) becomes:
\begin{equation}
\mathcal{A}^{\left(0\right)}_{n}\sum^{n}_{i=1}\left(z_{i}\frac{\partial}{\partial z_{i}}-\frac{1}{2}\omega_{i}\frac{\partial}{\partial\omega_{i}}\right)H_{n}=\frac{b^{2}}{4}\sum^{n}_{i=1}\left(\omega_{i}\frac{\partial}{\partial\omega_{i}}\right)^{2}\mathcal{A}^{\left(0\right)}_{n}.
\end{equation}

Since $\mathcal{A}^{\left(0\right)}_{n}$ represents the semiclassical
limit of the Liouville sector in our celestial CFT, we identify it
with the tree-level MHV $n$-gluon amplitude. From Eq. (\ref{eq:Parke-Taylor-3}):
\begin{equation}
\mathcal{A}^{\left(0\right)}_{n}=\frac{z^{4}_{12}}{z_{12}z_{23}...z_{n1}}\int\frac{d^{4}x}{\left(2\pi\right)^{4}}\prod^{n}_{i=1}\omega^{\alpha'_{i}}_{i}e^{i\omega_{i}q\left(z_{i},\bar{z}_{i}\right)\cdot x}.
\end{equation}
Then, we have:
\begin{equation}
\sum^{n}_{i=1}\left(\omega_{i}\frac{\partial}{\partial\omega_{i}}\right)^{2}\mathcal{A}^{\left(0\right)}_{n}=\sum^{n}_{i=1}\alpha^{2}_{i}\mathcal{A}^{\left(0\right)}_{n}=n\mathcal{A}^{\left(0\right)}_{n},
\end{equation}
and Eq. (\ref{eq:PDE-5}) reduces to:
\begin{equation}
\sum^{n}_{i=1}\left(z_{i}\frac{\partial}{\partial z_{i}}-\frac{1}{2}\omega_{i}\frac{\partial}{\partial\omega_{i}}\right)H_{n}=\frac{nb^{2}}{4}.
\end{equation}
The solution, respecting both global conformal symmetry and Lorentz
invariance, is:
\begin{equation}
H_{n}=-\frac{b^{2}}{4}\sum_{j=i+1}\log\frac{\omega_{i}\omega_{j}\left|z_{ij}\right|^{2}}{m^{2}},
\end{equation}
where $m^{2}$, with dimensions of energy squared, serves as the renormalisation
scale.

Consequently, the renormalised celestial Liouville amplitude for MHV
gluons is:
\begin{equation}
\mathcal{A}_{n}=\mathcal{R}_{n}A^{(0)}_{n}+\mathcal{O}(b^{4}),
\end{equation}
where the renormalisation factor is given by;
\begin{equation}
\mathcal{R}_{n}=1-\frac{b^{2}}{4}\sum_{j=i+1}\log\frac{\omega_{i}\omega_{j}\left|z_{ij}\right|^{2}}{m^{2}},\label{eq:Renormalisation-Factor-2}
\end{equation}
as defined in Eq. (\ref{eq:Renormalisation-Factor}) of Section \ref{sec:Celestial-OPE-Gravitons}.

The logarithmic correction should now be viewed as a finite-$b$ correction
to the full celestial Liouville theory amplitude. In the three-point
case it agrees with the one-loop correction constructed in Eq. (44)
of \citet{stieberger2023yang,stieberger2023celestial}.

\subsection{Correction to Celestial OPE for Gluons via Renormalised Liouville
Amplitudes}

In this subsection, we compute the correction to the celestial OPE
for gluons by incorporating contributions from the Liouville coupling
constant up to leading-order in $b^{2}$. Following a similar procedure
to the gravitational case, we utilise the collinear limit approach,
combined with the splitting function formalism as developed in \citet{bern1999multi,white2011factorization,akhoury2011collinear}.
In the context of celestial holography, we also draw upon the formalism
presented in \citet{pate2021celestial}. Ultimately, we demonstrate
that the deformation of the gluon celestial OPE, parametrised by $b^{2}$,
corresponds to the algebra obtained by considering the one-loop correction
to pure Yang-Mills theory.

In what follows, we consider the holomorphic collinear limit, where
the first two gluons become collinear, $z_{12}\rightarrow0$, while
$\bar{z}_{12}$ remains fixed. To simplify the exposition, we omit
anti-holomorphic variables from our expressions and suppress colour
indices until the conclusion of our calculations. According to \citet{bern1999multi},
\begin{align}
 & \lim_{z_{12}\rightarrow0}A_{n}\left(\omega_{1},z_{1};\omega_{2},z_{2};...;\omega_{n},z_{n}\right)\\
 & =\sum_{s=\pm1}\text{Split}^{s}_{s_{1}s_{2}}\left(p_{1},p_{2}\right)A_{n-1}\left(\omega_{1}+\omega_{2},z_{2};\omega_{3},z_{3};...;\omega_{n},z_{n}\right),
\end{align}
where $\text{Split}^{s}_{s_{1}s_{2}}\left(p_{1},p_{2}\right)$ is
the universal collinear factor, also known as the ``splitting function,''
for gluons. As described in \citet{elvang2013scattering,badger2024scattering},
the only helicity combinations yielding a non-vanishing splitting
function are:
\begin{equation}
\text{Split}^{1}_{11}\left(p_{1},p_{2}\right)=\frac{1}{z_{12}}\frac{\omega_{1}+\omega_{2}}{\omega_{1}\omega_{2}},
\end{equation}
and:
\begin{equation}
\text{Split}^{-1}_{1,-1}\left(p_{1},p_{2}\right)=\frac{1}{z_{12}}\frac{\omega_{2}}{\omega_{1}\left(\omega_{1}+\omega_{2}\right)}.
\end{equation}

We now focus on the case where both gluons have positive helicity,
$s_{1}=s_{2}=1$. Thus, we have:
\begin{align}
 & \lim_{z_{12}\rightarrow0}A_{n}\left(\omega_{1},z_{1};\omega_{2},z_{2};...;\omega_{n},z_{n}\right)\\
 & =\frac{1}{z_{12}}\frac{\omega_{1}+\omega_{2}}{\omega_{1}\omega_{2}}A_{n-1}\left(\omega_{1}+\omega_{2},z_{2};\omega_{3},z_{3};...;\omega_{n},z_{n}\right).\label{eq:Collinear-Limit-2}
\end{align}

Next, consider the renormalisation factor defined in Eq. (\ref{eq:Renormalisation-Factor-2}),
which can be written as:
\begin{align}
\mathcal{R}_{n}\left(\omega_{1},z_{1};\omega_{2},z_{2};...;\omega_{n},z_{n}\right) & =\mathcal{R}_{n-1}\left(\omega_{1}+\omega_{2},z_{2};\omega_{3},z_{3};...;\omega_{n},z_{n}\right)\\
 & +\frac{b^{2}}{2}\log\frac{\omega_{1}\omega_{2}\left|z_{12}\right|}{m\left(\omega_{1}+\omega_{2}\right)}.
\end{align}
This implies that the collinear limit of the renormalised celestial
Liouville amplitude is:
\begin{align}
 & \lim_{z_{12}\rightarrow0}\mathcal{A}_{n}\left(\omega_{1},z_{1};\omega_{2},z_{2};...;\omega_{n},z_{n}\right)\\
 & =\mathcal{R}_{n-1}\left(\omega_{1}+\omega_{2},z_{2};\omega_{3},z_{3};...;\omega_{n},z_{n}\right)\lim_{z_{12}\rightarrow0}A_{n}\left(\omega_{1},z_{1};\omega_{2},z_{2};...;\omega_{n},z_{n}\right)\\
 & +\frac{b^{2}}{2}\lim_{z_{12}\rightarrow0}\log\frac{\omega_{1}\omega_{2}\left|z_{12}\right|}{m\left(\omega_{1}+\omega_{2}\right)}A_{n}\left(\omega_{1},z_{1};\omega_{2},z_{2};...;\omega_{n},z_{n}\right)+\mathcal{O}\left(b^{2}\right).
\end{align}
Substituting Eq. (\ref{eq:Collinear-Limit-2}) into the above, and
retaining only terms up to leading order in $b^{2}$, we obtain:
\begin{align}
 & \lim_{z_{12}\rightarrow0}\mathcal{A}_{n}\left(\omega_{1},z_{1};\omega_{2},z_{2};...;\omega_{n},z_{n}\right)\\
 & =\frac{1}{z_{12}}\frac{\omega_{1}+\omega_{2}}{\omega_{1}\omega_{2}}\mathcal{A}_{n-1}\left(\omega_{1}+\omega_{2},z_{2};\omega_{3},z_{3};...;\omega_{n},z_{n}\right)\\
 & +\frac{b^{2}}{2}\frac{1}{z_{12}}\frac{\omega_{1}+\omega_{2}}{\omega_{1}\omega_{2}}\log\frac{\omega_{1}\omega_{2}\left|z_{12}\right|}{m\left(\omega_{1}+\omega_{2}\right)}\mathcal{A}_{n-1}\left(\omega_{1}+\omega_{2},z_{2};\omega_{3},z_{3};...;\omega_{n},z_{n}\right).
\end{align}

The collinear limit of the celestial amplitude is then obtained by
performing the Mellin transform of the above expression:
\begin{align}
 & \lim_{z_{12}\rightarrow0}\widehat{\mathcal{A}}_{n}\left(\omega_{1},z_{1};\omega_{2},z_{2};...;\omega_{n},z_{n}\right)=\\
 & \frac{1}{z_{12}}\left(\prod^{n}_{i=1}\int^{\infty}_{0}d\omega_{i}\omega^{\Delta_{i}-1}_{i}\right)\frac{\omega_{1}+\omega_{2}}{\omega_{1}\omega_{2}}\mathcal{A}_{n-1}\left(\omega_{1}+\omega_{2},z_{2};\omega_{3},z_{3};...;\omega_{n},z_{n}\right)\\
 & +\frac{b^{2}}{2}\frac{1}{z_{12}}\left(\prod^{n}_{i=1}\int^{\infty}_{0}d\omega_{i}\omega^{\Delta_{i}-1}_{i}\right)\frac{\omega_{1}+\omega_{2}}{\omega_{1}\omega_{2}}\log\frac{\omega_{1}\omega_{2}\left|z_{12}\right|}{m\left(\omega_{1}+\omega_{2}\right)}\mathcal{A}_{n-1}\left(\omega_{1}+\omega_{2},z_{2};...;\omega_{n},z_{n}\right).
\end{align}
To evaluate the Mellin integrals, we introduce new variables $t\in\left[0,1\right]$
and $\omega_{P}\in[0,\infty)$, defined by $\omega_{1}=t\omega_{P}$
and $\omega_{2}=\left(1-t\right)\omega_{P}$, such that:
\begin{equation}
d\omega_{1}d\omega_{2}\omega^{\Delta_{1}-1}_{1}\omega^{\Delta_{2}-1}_{2}\frac{\omega_{1}+\omega_{2}}{\omega_{1}\omega_{2}}=dtd\omega_{P}t^{\left(\Delta_{1}-1\right)-1}\left(1-t\right)^{\left(\Delta_{2}-1\right)-1}\omega^{\left(\Delta_{1}+\Delta_{2}\right)-1}_{P}.
\end{equation}
Utilising the following identity that relates the partial derivative
operator to the Mellin transform:
\begin{equation}
\frac{\partial}{\partial\Delta}\int^{\infty}_{0}d\omega\omega^{\Delta-1}f\left(\omega\right)=\int^{\infty}_{0}d\omega\omega^{\Delta-1}f\left(\omega\right)\log\omega,
\end{equation}
we find:
\begin{align}
 & \lim_{z_{12}\rightarrow0}\widehat{\mathcal{A}}_{n}\left(\omega_{1},z_{1};\omega_{2},z_{2};...;\omega_{n},z_{n}\right)=\\
 & \frac{1}{z_{12}}B\left(\Delta_{1}-1,\Delta_{2}-1\right)\widehat{\mathcal{A}}_{n-1}\left(\Delta_{1}+\Delta_{2}-1,z_{2};\Delta_{3},z_{3};...;\Delta_{n},z_{n}\right)\\
 & +\frac{b^{2}}{2}\frac{1}{z_{12}}\left(\frac{\partial}{\partial\Delta_{1}}+\frac{\partial}{\partial\Delta_{2}}+\frac{1}{2}\log\frac{\left|z_{12}\right|^{2}}{m^{2}}\right)B\left(\Delta_{1}-1,\Delta_{2}-1\right)\widehat{\mathcal{A}}_{n-1}\left(\Delta_{1}+\Delta_{2}-1,z_{2};...\right)\\
 & -\frac{b^{2}}{2}\frac{1}{z_{12}}B\left(\Delta_{1}-1,\Delta_{2}-1\right)\frac{\partial}{\partial\Delta}\bigg|_{\Delta=\Delta_{1}+\Delta_{2}-1}\widehat{\mathcal{A}}_{n-1}\left(\Delta,z_{2};\Delta_{3},z_{3};...;\Delta_{n},z_{n}\right).
\end{align}

Thus, we can now state the corrected celestial OPE for gluons, which
includes the leading correction arising from the Liouville coupling
parameter. Let $\mathcal{O}^{+a}_{\Delta}\left(z,\bar{z}\right)$
represent the gluon conformal primary of positive helicity with colour
index $a$. Reinstating the colour structure, we obtain the following
OPE:
\begin{align}
 & \mathcal{O}^{+a}_{\Delta_{1}}\left(z_{1},\bar{z}_{1}\right)\mathcal{O}^{+b}_{\Delta_{2}}\left(z_{2},\bar{z}_{2}\right)\\
 & \sim\frac{if^{abc}}{z_{12}}B\left(\Delta_{1}-1,\Delta_{2}-1\right)\mathcal{O}^{+c}_{\Delta_{1}+\Delta_{2}-1}\left(z_{2},\bar{z}_{2}\right)\\
 & +\frac{b^{2}}{2}\frac{1}{z_{12}}\left(\frac{\partial}{\partial\Delta_{1}}+\frac{\partial}{\partial\Delta_{2}}+\frac{1}{2}\log\frac{\left|z_{12}\right|^{2}}{m^{2}}\right)B\left(\Delta_{1}-1,\Delta_{2}-1\right)\mathcal{O}^{+c}_{\Delta_{1}+\Delta_{2}-1}\left(z_{2},\bar{z}_{2}\right)\\
 & -\frac{b^{2}}{2}\frac{1}{z_{12}}B\left(\Delta_{1}-1,\Delta_{2}-1\right)\frac{\partial}{\partial\Delta}\bigg|_{\Delta=\Delta_{1}+\Delta_{2}-1}\mathcal{O}^{+c}_{\Delta}\left(z_{2},\bar{z}_{2}\right).
\end{align}
An analogous calculation for the case of opposite helicity gluons,
where $\mathcal{O}^{-a}_{\Delta}\left(z,\bar{z}\right)$ denotes the
conformal primary representing a gluon with negative helicity, yields
the corresponding celestial OPE:
\begin{align}
 & \mathcal{O}^{+a}_{\Delta_{1}}\left(z_{1},\bar{z}_{1}\right)\mathcal{O}^{-b}_{\Delta_{2}}\left(z_{2},\bar{z}_{2}\right)\\
 & \sim\frac{if^{abc}}{z_{12}}B\left(\Delta_{1}-1,\Delta_{2}+1\right)\mathcal{O}^{-c}_{\Delta_{1}+\Delta_{2}-1}\left(z_{2},\bar{z}_{2}\right)\\
 & +\frac{b^{2}}{2}\frac{if^{abc}}{z_{12}}\left(\frac{\partial}{\partial\Delta_{1}}+\frac{\partial}{\partial\Delta_{2}}+\frac{1}{2}\log\frac{\left|z_{12}\right|^{2}}{m^{2}}\right)B\left(\Delta_{1}-1,\Delta_{2}+1\right)\mathcal{O}^{-c}_{\Delta_{1}+\Delta_{2}-1}\left(z_{2},\bar{z}_{2}\right)\\
 & -\frac{b^{2}}{2}\frac{if^{abc}}{z_{12}}B\left(\Delta_{1}-1,\Delta_{2}+1\right)\frac{\partial}{\partial\Delta}\bigg|_{\Delta=\Delta_{1}+\Delta_{2}-1}\mathcal{O}^{-c}_{\Delta}\left(z_{2},\bar{z}_{2}\right).
\end{align}

\section{Discussion\label{sec:Discussion}}

In these notes, we introduced the concept of \emph{celestial Liouville
amplitudes, }defined for MHV gravitons and gluons in Eqs. (\ref{eq:Celestial-Liouville-Amplitude})
and (\ref{eq:Celestial}). These amplitudes exhibit two important
properties. First, in the semiclassical limit $b\rightarrow0$, they
reduce to the tree-level scattering amplitudes for MHV gravitons and
gluons. Second, they are governed by a system of partial differential
equations parametrised by the Liouville coupling constant, whose leading-order
perturbative corrections in $b^{2}$ are logarithmic for both gravitons
and gluons. The corresponding celestial OPEs can be identified with
the one-loop corrections to the celestial OPEs of Yang-Mills and Einstein
gravity, as studied by \citet{bhardwaj2022loop,banerjee2024all,adamo2023all,krishna2024celestial,bittleston2023associativity}.
This indicates that celestial Liouville theory encodes the one-loop
corrections to gauge theory and perturbative gravity in flat space.

It is intriguing, from a heuristic perspective, that the study of
celestial holography, originally based in the investigation of infrared
(IR) physics and its relation to asymptotic symmetries, has evolved
into the development of toy models for celestial CFTs that may describe
some aspects of the ultraviolet (UV) regime of realistic theories
in flat spacetime backgrounds. This development presents a subtle
tension, though not a strict logical inconsistency, with the foundational
principle of the Wilsonian paradigm in effective field theory, as
reviewed by \citet{weinberg2016effective,weinberg2021development},
since in the Wilsonian framework, one would typically expect a decoupling
of IR and UV degrees of freedom. A comprehensive discussion expressing
similar views is provided by \citet{arkani2021celestial}.

A natural direction for extending the results of this work is to explore
the extent to which celestial Liouville theories can account for loop
corrections in gauge theories. Additionally, it would be interesting
to investigate whether an analogous construction could be found for
Carrollian amplitudes, as review by \citet{alday2024carrollian,mason2024carrollian}.
Specifically, one might ask: Is there a toy model CFT that holographically
reproduces Carrollian amplitudes? We believe the answer is affirmative,
with a hint provided by \citet{bagchi2023ads}, where Carrollian correlators
are expressed as $AdS$ Feynman-Witten diagrams, in a manner reminiscent
of Eqs. (\ref{eq:Step}, \ref{eq:Step-1}, \ref{eq:Step-2}), which
allowed us to interpret the semiclassical limit of Liouville correlation
functions as contact Feynman-Witten diagrams for massless scalars
propagating in $AdS_{3}$.

Moreover, we note that the Parke-Taylor factor in Eqs. (\ref{eq:Celestial-Liouville-Amplitude})
and (\ref{eq:Celestial}) may be interpreted as the large-$N$ limit
of level-one $SO\left(2N\right)$ $WZNW$currents, $J^{a}$, on the
celestial sphere, as detailed in \ref{subsec:Celestial-Amplitude-for},
Eqs. (\ref{eq:WZNW}, \ref{eq:Celestial-Liouville-Amplitude}). An
intriguing question arises as to whether non-MHV amplitudes could
be derived by considering $WZNW$ models in the context of celestial
amplitudes. Could there exist a (potentially dressed) Kac-Moody current
algebra on the celestial sphere capable of holographically generating
non-MHV amplitudes for both gravitons and gluons? If so, this approach
may offer a new perspective on the celestial stress tensor through
the Sugawara construction\footnote{I thank Shamik Banerjee for bringing this point to my attention.}.

\section{Acknowledgement}

This research was supported in part by Perimeter Institute for Theoretical
Physics during the ``Simons/Perimeter Celestial Holography Summer
School 2024.'' Research at Perimeter Institute is supported by the
Government of Canada through the Department of Innovation, Science
and Economic Development and by the Province of Ontario through the
Ministry of Research, Innovation and Science. The author also thanks
Giorgio Torrieri for his important encouragement.

\appendix

\section{Operator Decomposition\label{sec:Operator-Factorisation}}

In this Appendix, we shall establish the following mathematical theorem,
which is important in analysing the factorisation of the $n$-graviton
celestial amplitude into linear differential operators:

Let $\nu^{A}_{2},\nu^{A}_{3},...,\nu^{A}_{n}\in\mathbf{C}^{2}$ and
$\bar{\nu}^{\dot{A}}_{2},\bar{\nu}^{\dot{A}}_{3},...,\bar{\nu}^{\dot{A}}_{n}\in\mathbf{C}^{2}$
be sequences of two-component spinors, and let $z_{2},z_{3},...,z_{n}\in\mathbf{CP}^{1}$
and $\bar{z}_{2},\bar{z}_{3},...,\bar{z}_{n}\in\mathbf{CP}^{1}$ represent
their respective local coordinates on an open neighbourhood of the
complex projective line. These satisfy $\left\langle \nu_{i},\nu_{j}\right\rangle \coloneqq\varepsilon_{AB}\nu^{A}_{i}\nu^{B}_{j}=z_{ij}$
and $\left\langle \bar{\nu}_{i},\bar{\nu}_{j}\right\rangle \coloneqq\varepsilon_{\dot{A}\dot{B}}\bar{\nu}^{\dot{A}}_{i}\bar{\nu}^{\dot{B}}_{j}=\bar{z}_{ij}$
for each pair of indices $1\leq i,j\leq n$. 

Furthermore, let $\mathsf{A}_{i}$ and $\mathsf{B}_{i}$ be linear
differential operators with respect to the real variables $u_{1},u_{2},...,u_{n}\in\mathbb{R}$
defined by\footnote{It is important to note that the operators $\mathsf{A}_{i}$ differs
from the ones defined in Eq. (\ref{eq:A-Operator}). The latter operators
contain the factor $\chi^{\dagger}\left(z\right)\chi\left(z\right)$,
where $\left(\chi,\chi^{\dagger}\right)$ is a doublet of free quasi-primary
fermionic fields residing on $\mathbf{CP}^{1}$ defined in Section
\ref{sec:Gravitons}. This factor is omitted here as it is not pertinent
to the current discussion.}:
\begin{equation}
\mathsf{A}_{i}\coloneqq e^{q\left(z_{i},\bar{z}_{i}\right)\cdot y\frac{\partial}{\partial u_{i}}},\,\,\,\mathsf{B}_{i}\coloneqq\frac{\bar{\nu}^{\dot{A}}_{i}\lambda^{A}}{\left\langle \nu_{i},\lambda\right\rangle }\frac{\partial}{\partial y^{A\dot{A}}}e^{q\left(z_{i},\bar{z}_{i}\right)\cdot y\frac{\partial}{\partial u_{i}}},\label{eq:Definitions}
\end{equation}
where $\lambda^{A}\in\mathbf{C}^{2}$ is a fixed auxiliary two-component
spinor, and $\lambda\in\mathbf{CP}^{1}$ denotes its local coordinate
on the complex projective line. The four-vector $y^{\mu}$ should
be viewed as a continuous parameter labelling these operators.

The central assertion we put forward is the following identity concerning
the operator expansion of a sequence of $\mathsf{B}_{i}$ operators
acting on the product $\mathsf{A}_{n-1}\mathsf{A}_{n}$: 
\begin{equation}
\left(\prod^{n-2}_{k=2}\mathsf{B}_{k}\right)\mathsf{A}_{n-1}\mathsf{A}_{n}=\prod^{n}_{r=2}e^{q\left(z_{r},\bar{z}_{r}\right)\cdot y\frac{\partial}{\partial u_{r}}}\left(\prod^{n-2}_{i=2}\sum^{n}_{j_{i}=i+1}\right)\prod^{n-2}_{k=2}\frac{\left(\bar{z}_{k}-\bar{z}_{j_{k}}\right)\left(z_{j_{k}}-\lambda\right)}{z_{k}-\lambda}\frac{\partial}{\partial u_{j_{k}}}.\label{eq:Proposition}
\end{equation}

Using the definitions provided in Eq. (\ref{eq:Definitions}), it
is instructive to calculate the first instances of Eq. (\ref{eq:Proposition}).
This not only aids in developing an intuition for the formula but
also serves to justify our induction hypothesis. 

First, we have:
\begin{align}
\mathsf{B}_{2}\mathsf{A}_{3}\mathsf{A}_{4} & =\prod^{4}_{r=2}e^{q\left(z_{r},\bar{z}_{r}\right)\cdot y\frac{\partial}{\partial u_{r}}}\left(\frac{\left(\bar{z}_{2}-\bar{z}_{3}\right)\left(z_{3}-\lambda\right)}{z_{2}-\lambda}\frac{\partial}{\partial u_{3}}+\frac{\left(\bar{z}_{2}-\bar{z}_{4}\right)\left(z_{4}-\lambda\right)}{z_{2}-\lambda}\frac{\partial}{\partial u_{4}}\right)\\
 & =\prod^{4}_{r=2}e^{q\left(z_{r},\bar{z}_{r}\right)\cdot y\frac{\partial}{\partial u_{r}}}\sum^{4}_{j_{2}=2+1}\prod^{4-2}_{k=2}\frac{\left(\bar{z}_{k}-\bar{z}_{j_{k}}\right)\left(z_{j_{k}}-\lambda\right)}{z_{k}-\lambda}\frac{\partial}{\partial u_{j_{k}}},
\end{align}

Second, for the case $n=5$, we compute:
\begin{align}
 & \mathsf{B}_{2}\mathsf{B}_{3}\mathsf{A}_{4}\mathsf{A}_{5}=\prod^{5}_{i=2}e^{q\left(z_{i},\bar{z}_{i}\right)\cdot y\frac{\partial}{\partial u_{i}}}\frac{\bar{\nu}^{\dot{A}}_{2}\lambda^{A}}{\left\langle \nu_{2},\lambda\right\rangle }\frac{\bar{\nu}^{\dot{B}}_{3}\lambda^{B}}{\left\langle \nu_{3},\lambda\right\rangle }\\
 & \bigg(\nu_{3A}\bar{\nu}_{3\dot{A}}\frac{\partial}{\partial u_{3}}+\nu_{4A}\bar{\nu}_{4\dot{A}}\frac{\partial}{\partial u_{4}}+\nu_{5A}\bar{\nu}_{5\dot{A}}\frac{\partial}{\partial u_{5}}\bigg)\bigg(\nu_{4B}\bar{\nu}_{4\dot{B}}\frac{\partial}{\partial u_{4}}+\nu_{5B}\bar{\nu}_{5\dot{B}}\frac{\partial}{\partial u_{5}}\bigg),
\end{align}
which simplifies to:
\begin{equation}
\mathsf{B}_{2}\mathsf{B}_{3}\mathsf{A}_{4}\mathsf{A}_{5}=\prod^{5}_{r=2}e^{q\left(z_{r},\bar{z}_{r}\right)\cdot y\frac{\partial}{\partial u_{r}}}\sum^{5}_{j_{2}=2+1}\sum^{5}_{j_{3}=3+1}\prod^{5-2}_{k=2}\frac{\left(\bar{z}_{k}-\bar{z}_{j_{k}}\right)\left(z_{j_{k}}-\lambda\right)}{z_{k}-\lambda}\frac{\partial}{\partial u_{j_{k}}}.
\end{equation}

Thus, Eq. (\ref{eq:Proposition}) is seen to correctly reproduce the
results for $n=4$ and $n=5$, respectively. This observation leads
us to posit the following induction hypothesis:\emph{ there exists
an integer $N>3$ such that }Eq. (\ref{eq:Proposition})\emph{ holds
for $n=N$. }

Now, consider the following computation:
\begin{align}
 & \left(\prod^{n-2}_{i=1}\mathsf{B}_{i}\right)\mathsf{A}_{n-1}\mathsf{A}_{n}=\frac{\bar{\nu}^{\dot{A}}_{1}\lambda^{A}}{\left\langle \nu_{1},\lambda\right\rangle }\frac{\partial}{\partial y^{A\dot{A}}}e^{q\left(z_{1},\bar{z}_{1}\right)\cdot y\frac{\partial}{\partial u_{1}}}\\
 & \Bigg[\prod^{n}_{r=2}e^{q\left(z_{r},\bar{z}_{r}\right)\cdot y\frac{\partial}{\partial u_{r}}}\left(\prod^{n-2}_{i=2}\sum^{n}_{j_{i}=i+1}\right)\prod^{n-2}_{k=2}\frac{\left(\bar{z}_{k}-\bar{z}_{j_{k}}\right)\left(z_{j_{k}}-\lambda\right)}{z_{k}-\lambda}\frac{\partial}{\partial u_{j_{k}}}\Bigg]\\
 & =\frac{\bar{\nu}^{\dot{A}}_{1}\lambda^{A}}{\left\langle \nu_{1},\lambda\right\rangle }\frac{\partial}{\partial y^{A\dot{A}}}\Bigg[\prod^{n}_{r=1}e^{q\left(z_{r},\bar{z}_{r}\right)\cdot y\frac{\partial}{\partial u_{r}}}\left(\prod^{n-2}_{i=2}\sum^{n}_{j_{i}=i+1}\right)\prod^{n-2}_{k=2}\frac{\left(\bar{z}_{k}-\bar{z}_{j_{k}}\right)\left(z_{j_{k}}-\lambda\right)}{z_{k}-\lambda}\frac{\partial}{\partial u_{j_{k}}}\Bigg]\\
 & =\prod^{n}_{r=1}e^{q\left(z_{r},\bar{z}_{r}\right)\cdot y\frac{\partial}{\partial u_{r}}}\left(\prod^{n-2}_{i=1}\sum^{n}_{j_{i}=i+1}\right)\prod^{n-2}_{k=2}\frac{\left(\bar{z}_{k}-\bar{z}_{j_{k}}\right)\left(z_{j_{k}}-\lambda\right)}{z_{k}-\lambda}\frac{\partial}{\partial u_{j_{k}}},\label{eq:Last-Line}
\end{align}

The last line of Eq. (\ref{eq:Last-Line}) follows from an iterative
application of the Leibniz rule for the differential operator $\partial/\partial y^{A\dot{A}}$.
Therefore, if Eq. (\ref{eq:Proposition}) holds for $n=N$, it also
holds for $n=N+1$, completing our inductive argument. \textbf{QED.}

\bibliographystyle{revtex-tds/bibtex/bst/revtex/aipnum4-2}
\bibliography{CCFT2}

\end{document}